\documentclass[useAMS,fleqn,usenatbib]{mnras}
\usepackage[utf8]{inputenc}
\usepackage{amssymb,amsmath}
\usepackage{graphicx}
\usepackage[dvipsnames]{xcolor}

\usepackage{newtxtext,newtxmath}

\usepackage{array}
\newcolumntype{P}[1]{>{\centering\arraybackslash}p{#1}}

\newcommand{\eqn}[1]{equation (\ref{#1})}

\newcommand{\fig}[1]{Figure \ref{#1}}
\newcommand{\de}{\mathrm{d}}

\title[\texttt{CosmoReionMC}: parameter estimation using CMB and reionization data]{\texttt{CosmoReionMC}: A package for estimating cosmological and astrophysical parameters using CMB, Lyman-$\alpha$ absorption and global 21~cm data}

\author[Chatterjee, Choudhury \& Mitra]{
Atrideb Chatterjee$^{1}$\thanks{E-mail:atrideb@ncra.tifr.res.in},
Tirthankar Roy Choudhury$^{1}$\thanks{E-mail:tirth@ncra.tifr.res.in}
and
Sourav Mitra$^{2}$
\\
$^{1}$National Centre for Radio Astrophysics, TIFR, Post Bag 3, Ganeshkhind, Pune 411007, India\\
$^{2}$ Department of Physics, Surendranath College, 24/2 M. G. Road, Kolkata 700009, India
}

\date{Accepted XXX. Received YYY; in original form ZZZ}

\pubyear{2020}

\begin{document}
\label{firstpage}
\pagerange{\pageref{firstpage}--\pageref{lastpage}}
\maketitle

\begin{abstract}
We present a Markov Chain Monte Carlo (MCMC)-based parameter estimation package, \texttt{CosmoReionMC}, to jointly constrain cosmological parameters of the $\Lambda$CDM model and the astrophysical parameters related to hydrogen reionization.
The package is based on a previously developed physically motivated semi-analytical model for reionization, a similar semi-analytical model for computing the global 21~cm signal during the cosmic dawn and using an appropriately modified version of the publicly available CAMB for computing the CMB anisotropies.
These calculations are then coupled to an MCMC ensemble sampler \texttt{emcee} to compute the posterior distributions of the model parameter.
The model has twelve free parameters in total: five cosmological and seven related to the stellar populations.
We constrain the parameters by matching the theoretical predictions with CMB data from Planck, observations related to the quasar absorption spectra and, for the first time, the global 21~cm signal from EDGES.
We find that incorporating the quasar spectra data in the analysis tightens the bounds on the electron scattering optical depth $\tau$ and consequently the normalization $A_s$ of the primordial matter power spectrum (or equivalently $\sigma_8$).
Furthermore, when we include the EDGES data in the analysis, we find that an early population of metal-free stars with efficient radio emission is necessary to match the absorption amplitude.
The \texttt{CosmoReionMC} package should have interesting future applications, e.g., probing non-standard extensions to the $\Lambda$CDM model.
\end{abstract}

\begin{keywords}
intergalactic medium -- 
dark ages, reionization, first stars --
stars: Population III --
cosmology: theory.
\end{keywords}



\section{Introduction}

Our Universe has gone through two-phase transitions between the ionized and neutral states of the hydrogen atom, one at $z \approx 1100$, when the free electrons from ionized plasma recombine with the protons to form atomic hydrogen and another at $z \approx 6$ when the neutral Universe becomes almost completely ionized. The later epoch $(z \approx 6-20)$, when the first stars form and the photons coming from them, start to heat up and ionize the intergalactic medium (IGM), is broadly known as Epoch of Reionization (EoR) \citep{2001PhR...349..125B, 2001ARA&A..39...19L, 2009CSci...97..841C, 2018PhR...780....1D}. The very early phase of reionization when the first stars just started to form, is sometimes referred to as the Cosmic Dawn (CD). The Cosmic Microwave Background (CMB) photons coming from the last scattering surface of the Universe (at $z \approx 1100$) interact with the free electrons in the IGM produced during the EoR. As a result of this interaction, the CMB temperature anisotropy gets suppressed whereas the amplitude of the EE polarization of the CMB anisotropy angular power spectrum gets enhanced \citep{1980MNRAS.190..413S, 1987ApJ...322..597V, 2000ApJ...529...12H, 2002ARA&A..40..171H, 2005ApJ...630..643M}. It is well known that using the observed CMB anisotropies and assuming a reionization history, the direction-averaged Thomson scattering optical depth, $\tau$, can be computed. It has been a common practice to compare the value of the optical depth estimated from the reionization model with that of the CMB angular power spectrum $C_l$ observations to constrain the free parameters of the model \citep{2016A&A...594A..13P, 2020A&A...641A...6P, 2003PhRvD..68b3001H, 2012MNRAS.419.1480M, 2008ApJ...672..737M, 2011PhRvD..84l3522P, 2015MNRAS.454L..76M, 2018MNRAS.473.1416M}.

While calculating $\tau$ from the observed CMB angular power spectra,  one usually assumes a redshift-symmetric tanh model of reionization \citep{2008PhRvD..78b3002L, 2016A&A...594A..13P, 2020A&A...641A...6P} or a model described by polynomials \citep{2017JCAP...11..028H}. However, the value of optical depth, in principle, depends on the details of the reionization history used, and hence the optical depths calculated from the tanh or polynomial models should not be compared with that estimated from the model which assumes a different reionization history. One way to overcome the problem is to connect a reionization model with a ``CMB anisotropy code'' e.g. Code for Anisotropies in the Microwave Background (CAMB) \citep{Lewis:2013hha} or the Cosmic Linear Anisotropy Solving System (CLASS) \citep{2011arXiv1104.2932L} and then compare with the observed CMB power spectrum to constrain the free parameters of the reionization model \citep{2011PhRvD..84l3522P, 2012MNRAS.419.1480M}. Such analysis with a physically motivated reionization model had been carried out in the past \citep{2017MNRAS.467.4050M, 2020MNRAS.tmp.2712Q}. However, in \cite{2017MNRAS.467.4050M}, the authors varied only the model parameters and $\tau$ but not any other cosmological parameters, whereas other studies \citep{2017MNRAS.467.4050M, 2020A&A...641A...6P, 2015A&A...580L...4D, 2015MNRAS.454L..76M} suggest that parameters related to reionization have degeneracies with the cosmological parameters and therefore could possibly introduce bias in the obtained constraints on the reionization model parameters. Hence the right approach to handle this problem would be to constrain the cosmological and the reionization parameters simultaneously \citep{2012MNRAS.419.1480M, 2020MNRAS.tmp.2712Q,2020PhRvL.125g1301H} by comparing with CMB and astrophysical data sets.

Recently a sky-averaged global 21~cm signal spanning the redshift range $z \approx 13-22$ has been claimed to be detected by Experiment to Detect the Global EoR Signal (EDGES) collaboration \citep{2018Natur.555...67B}. Although the ``true" origin of the signal is still in debate \citep{2018Natur.564E..32H, 2019ApJ...874..153B, 2019ApJ...880...26S, 2020MNRAS.492...22S}, it is indeed a promising probe to find out the epoch of structure formation (for review look at \citealt{2006PhR...433..181F}) as this signal captures the information about first stars during the cosmic dawn \citep{2020MNRAS.496.1445C, 2020MNRAS.493.1217M, 2019ApJ...877L...5S}. However, modelling of the 21~cm signal at the cosmic dawn depends on the cosmological parameters. For example, any parameter which affects the formation of dark matter haloes (say, the amplitude of the primordial matter power spectrum) will also affect the timing of the formation of first stars and shift the absorption trough of this signal to a different redshift. Thus, adding global 21~cm observation along with the reionization and CMB observations should potentially provide an independent way to put new constraints on the cosmological parameters. In fact, previous studies \citep{2016PhRvD..93d3013L} had predicted that the degeneracy between $\tau$ and $A_{s}$ can be allevated by using the Global 21~cm signal. But of course at that time the global 21~cm signal had not been detected so they could not apply their method on the actual observation.

A huge body of works in recent years \citep{2015MNRAS.449.4246G, 2016MNRAS.459.2342M, 2017MNRAS.465.4838G,2017MNRAS.472.2651G, 2017MNRAS.468..122H, 2017ApJ...848...23K, 2018MNRAS.477.3217G,2018MNRAS.475.1213S, 2019MNRAS.484..933P, 2019MNRAS.484..282G, 2021MNRAS.503.4551G} have been devoted in building different kinds of parameter estimation techniques which use 21~cm signal (both global and power spectrum) and reionization related observations to put constraints on the model parameters. However, they varied only the model parameters and kept the cosmological parameters fixed to their best fit values, except \cite{2017ApJ...848...23K} varied both the model parameters and the cosmological parameters. In \cite{2017ApJ...848...23K}, the authors also varied the cosmological parameters along with the model parameters and they use an emulator based MCMC approach to constrain these parameters using 21~cm power spectrum as their observation. 
 
 In this work, we develop a parameter estimation routine, called \texttt{CosmoReionMC}, which combines a semi-analytical physically motivated reionization and cosmic dawn model with the publicly available Boltzmann solver code CAMB and perform a Markov Chain Monte Carlo (MCMC) analysis to put constraints on the cosmological and model parameters simultaneously using the CMB, reionization and global 21~cm (EDGES) observation. The aim of this work is two-fold: First, we investigate in detail the impact of choosing a physically motivated reionization scenario over the tanh model on the constraints on the $\Lambda$CDM cosmological parameters with the most recent CMB data, thus updating the results obtained by \citet{2015MNRAS.454L..76M, 2018MNRAS.473.1416M}. Secondly, for the first time, we explore the effect of combining the global 21~cm experiments with the reionization and CMB observations (with full likelihoods of the CMB data) on the parameter estimation.

The paper is organized as follows. In Section \ref{section:theoretical_model}, we describe the theoretical model used to compute the reionization and 21~cm signal. Section \ref{section:reionization_CMB} describes the joint constraints on cosmological and reionization parameters after adding reionization observations with CMB observations. The effect of including the 21~cm observation along with the reionization and CMB observation is discussed in detail in Section \ref{section:CMB_reionization_21}. We summarize our results and conclusion in Section \ref{section:conclusion}.

\section{The Theoretical Model}
\label{section:theoretical_model}

The theoretical model used in this work contains several components which have been combined efficiently to compute the observable. We discuss the model components one by one.

\subsection{Reionization Model}

We use a reionization model proposed by \cite{2005MNRAS.361..577C}, \cite{2006MNRAS.371L..55C} (referred as CF model hereafter) and later slightly modified in \cite{2011MNRAS.413.1569M}. Here we summarize the main features of the reionization model used in this analysis.
\begin{enumerate}
    
\item The model computes the evolution of the ionization and thermal state of the neutral, HII and HeIII regions of the IGM separately, simultaneously and self-consistently.
    
\item The inhomogeneities of the intergalactic medium (IGM) is described using the method outlined in \cite{2003ApJ...597...66M}. The probability distribution of the over-densities are assumed to follow a log-normal distribution in the low density region, whereas the high density region of the IGM is described using a power law distribution. In this model, the reionization is assumed to be complete once all the low density regions are ionized, i.e., the volume filling factor $Q_{\mathrm{HII}}$ of ionized hydrogen in low density regions becomes unity. 
    
\item The sources of reionization in this model are assumed to be quasars, PopII and PopIII stars. For the stellar sources, the photon production rate is assumed to be of the form
\begin{align}
    \dot{n}_{\nu}(z) &= \dot{n}_{\nu}^{\mathrm{II}} + \dot{n}_{\nu}^{\mathrm{III}}
    \notag \\
    &= \rho_b \left[
    f_*^{\mathrm{II}} f_{\mathrm{esc}}^{\mathrm{II}} \left(\frac{\de N_{\nu}}{\de M} \right)_{\mathrm{II}} \frac{\de f_{\mathrm{coll, II}}}{\de t}
    + f_*^{\mathrm{III}} f_{\mathrm{esc}}^{\mathrm{III}} \left(\frac{\de N_{\nu}}{\de M} \right)_{\mathrm{III}} \frac{\de f_{\mathrm{coll, III}}}{\de t}
    \right],
    \label{eq:ndot_II_III}
\end{align}
where $\rho_b$ is the mean comoving density of baryons in the IGM, $f_{*}$ is the star formation efficiency, $f_{\mathrm{esc}}$ is the escape fraction of the ionizing photons entering IGM and $\de N_{\nu} / \de M$ is the number of photons emitted per frequency range per unit mass of stars. The total number of hydrogen ionizing photons from stellar sources is then
\begin{align}
        \dot{n}_{\mathrm{ph}}(z) &= \int_{\nu_H}^{\infty} \de \nu~\dot{n}_{\nu}(z)
        \notag \\
        &= n_{b}\left[N_{\mathrm{ion, II}}\left(\frac{\de f_{\mathrm{coll, II}}}{\de t}\right)+  N_{\mathrm{ion,III}}\left(\frac{\de f_{\mathrm{coll, III}}}{\de t}\right)\right],
    \end{align} 
where $\nu_H$ is the threshold frequency for hydrogen photoionization, $n_{b}$ is the mean comoving number density of baryons in the IGM and 
\begin{equation}
   N_{\mathrm{ion}} \equiv m_{p}\int^{\infty}_{\nu_H} f_{*} f_{\mathrm{esc}} \left(\frac{\de N_{\nu}}{\de M} \right) \de \nu
   \label{eq:Nion}
\end{equation}
is the number of ionizing photons escaping into the IGM per baryons in stars, with $m_p$ being the proton mass. In this work, we assume the parameters $f_*, f_{\mathrm{esc}}$ for each stellar population to be independent of the halo mass $M$ and redshift $z$ which allows us to keep the number of free parameters under control. These assumptions are clearly simplified as there are indications that these have non-trivial dependencies on $M$ and $z$, however, there is still no consensus on the actual form for this dependence \citep[see,e.g.,][]{2012MNRAS.427.2889Y,2013MNRAS.428L...1M,2014MNRAS.442.2560W, 2015MNRAS.451.2544P, 2017MNRAS.466.4826K, 2019MNRAS.486.2215K, 2017MNRAS.472.2009Q, 2019MNRAS.489..977D, 2020MNRAS.491.1398G, 2020MNRAS.491.3891P}.

The fraction of mass in collapsed haloes forming PopII and PopIII stars at redshift $z$ are given by
    \begin{equation}
        f_{\mathrm{coll,II}} = \frac{1}{\rho_{m}}\int^{\infty}_{M_{\mathrm{min}}(z)} \de M \left[1 - f_{\mathrm{III}}(M,z) \right] M \frac{\partial n(M,z)}{\partial M},
    \end{equation}
and
    \begin{equation}
        f_{\mathrm{coll,III}} = \frac{1}{\rho_{m}}\int^{\infty}_{M_{\mathrm{min}}(z)} \de M f_{\mathrm{III}}(M,z) M \frac{\partial n(M,z)}{\partial M},
    \end{equation}
    where $\rho_{m}$ is the mean comoving density of dark matter and $\partial n/ \partial M$ is number density of haloes per unit comoving volume per unit mass range, computed using the Sheth-Tormen form \citep{1999MNRAS.308..119S, 2001MNRAS.323....1S}. The quantity $f_{\mathrm{III}}(M,z)$ is the fraction of haloes of mass $M$ that are forming PopIII stars at $z$.

A crucial ingredient of the model is to compute the transition from metal-free PopIII to the usual PopII stars, which essentially determines the form of $f_{\mathrm{III}}(M,z)$. The transition is regulated by chemical feedback from the PopIII stars and hence depends on the merger history of dark matter haloes and mixing of metals in the IGM \citep{jaacks2018, sarmento2019, maio2010, maio2011}. In reality, the transition is likely to be complex with different regions of the universe being metal-contaminated at different times. Modelling the effect in a self-consistent manner is beyond the scope of any semi-analytical model like ours. We thus assume that the PopIII $\longrightarrow$ PopII transition occurs at a redshift $z_{\mathrm{trans}}$ with a width $\delta z$, the exact form being given by 
\begin{equation}
  f_{\mathrm{III}}(M,z) = f_{\mathrm{III}}(z)=\frac{1}{2} \left[1 + \tanh \left(\frac{z-z_{\mathrm{trans}}}{\delta z} \right) \right].
\end{equation}
The transition redshift $z_{\mathrm{trans}}$ is a free parameter of the model while the width $\delta z$ is taken to be that corresponding to the dynamical time-scale of collapsed objects at $z_{\mathrm{trans}}$. The reason for choosing so is because the star formation in haloes is expected to be spread over the dynamical time \citep{2000ApJ...534..507C}, hence the same time-scale will determine the decline of the PopIII stars. Furthermore, we find that typically $\delta z \sim 2$ at $z_{\mathrm{trans}} \sim 16$ which is consistent with the parametric PopIII star formation rate found by \citet{2020MNRAS.496.1445C} to match the EDGES data.

As discussed in \cite{2005MNRAS.361..577C}, the quantity $\de N_{\nu}/ \de M$ depends on the stellar spectra and IMF of the stars. For PopII stars, this number has been calculated from the stellar population synthesis models of \citet{2003MNRAS.344.1000B} using a standard Salpeter IMF in the mass range $1-100 M_{\odot}$ with a metallicity of $0.05 M_{\odot}$. For PopIII stars, it is calculated using stellar spectra of very high mass stars $(>300 M_{\odot})$ \citep{2002A&A...382...28S}.

We further fix the PopII star formation efficiency $f^{\mathrm{II}}_{*}$ by matching our predictions with the observed UV luminosity function ($6 \le z\le 10$) \citep{2006ApJ...653...53B, 2015ApJ...803...34B, 2018ApJ...855..105O, 2014ApJ...786..108O, 2017ApJ...835..113L, 2018ApJ...854...73I}. Obviously, we implicitly assume here that the observed stars at these redshifts are dominantly PopII. This is justified because, as we will see later, the data implies the PopIII stars to decline by $z \sim 15$. We found that to match with the observed UV luminosity function, we require $f^{\mathrm{II}}_{*} \sim 0.01$, and hence we fix this value throughout the paper.\footnote{The values of $f^{\mathrm{II}}_{*}$ in this work are lower than those obtained by \citet{2015MNRAS.454L..76M}. The main reason for this is that \citet{2015MNRAS.454L..76M} used the halo mass function of \citet{1974ApJ...187..425P}, while we are using the one given by \citep{1999MNRAS.308..119S, 2001MNRAS.323....1S}. In addition, the UV luminosity function data have been updated to include all the latest observations.}

For the quasars, the rate of ionizing photons per unit volume per unit frequency range has been computed following the procedure outlined in \citet{2005MNRAS.361..577C}, i.e., from the relation
\begin{equation}
    \label{eq:equation_6}
    \dot{n}_{\nu,Q}(z)=\int^{\infty}_{0} \de L_{B}~L_{B}~\psi(L_{B},z) \frac{L_{\nu}(L_{B})}{h_p \nu},
\end{equation}
where $\psi$ is the quasar luminosity function taken from the recent compilation of \citet{2019MNRAS.488.1035K}, $L_{B}$ is the B-band luminosity and $h_{p}$ is the Planck's constant. Assuming the luminosity $(L_{\nu})$ of all quasars to follow a power law with $L_{\nu} \propto \nu^{-\alpha}$ with $\alpha=1.57$ for $\lambda \sim 500$ to $\sim 1200$\AA\  \citep{2003ApJ...584..110S, 2002ApJ...565..773T},  we can write
\begin{equation}
    \label{eq:equation_7}
    \frac{L_{\nu}(L_{B})}{\mathrm{erg}\, \mathrm{s}^{-1}\, \mathrm{Hz}^{-1}}=\frac{L_{B}}{L_{\odot,B}} 10^{18.05} \left(\frac{\nu}{\nu_{H}}\right)^{-1.57},
\end{equation}
where $\nu_H$ is the threshold frequency for photoionization of hydrogen. Putting \eqn{eq:equation_7} in \eqn{eq:equation_6} one can calculate the ionizing photon production rate from quasars. The only role the quasars play in our model is to reionize HeII at $z \sim 3.5$. Since most of the work in this paper is concentrated on the ionization of hydrogen, the details of the ionizing photons from quasars can be ignored for the rest of the paper.

\item The value of the collapsed fraction $f_{\mathrm{coll}}$ depends on the minimum mass $M_{\mathrm{min}}(z)$ of haloes that are capable of forming stars and producing ionizing photons. In the neutral regions, we consider only atomically cooled haloes while calculating the minimum mass.\footnote{There are studies which indicate that the first stars would have formed in minihaloes that are cooled via H$_2$ molecules \citep{1997ApJ...476..458H, 2003Natur.425..812B, 2016MNRAS.457.3356V} . The efficiency of these haloes will depend on the severity of the dissociating Lyman-Werner (LW) feedback \citep{2001ApJ...561L..55O, 2010MNRAS.402.1249S, 2011MNRAS.418..838W, 2013MNRAS.432.2909F, 2014MNRAS.445..107V}. The presence of molecular cooling and LW feedback have been ignored in this work.} In addition, the radiative feedback from stars has also been implemented to account for the fact that once the first stars form they will ionize and heat the surrounding IGM. This implies that the minimum halo mass for star formation is higher in the ionized regions \citep{2016MNRAS.463.1462O,2019A&A...626A..77O, 2019MNRAS.483.1029K, 2019MNRAS.490.3177W, 2020arXiv200408401H} and thus star formation is suppressed in low mass haloes through a Jeans mass prescription \citep{2005MNRAS.361..577C}. As our model tracks the thermal history of the ionized region separately from the neutral region of the IGM, this effect has easily been implemented while computing the minimum halo mass of the star forming region.

\item Assuming that only the high density regions of the IGM determine the mean free path of the photons \citep[as discussed in][]{2003ApJ...597...66M}, the CF reionization model computes it as
\begin{equation}
\label{eq:equation_3}
    \lambda_{\mathrm{mfp}}=\frac{\lambda_{0}}{[1-F_{v}(\Delta_{i})]^{2/3}}
\end{equation}
 where $\lambda_{0}$ is a free parameter of the model and $F_{V}$ is the volume fraction of the ionized region, i.e., regions with overdensities less than $\Delta_i$ (the high-density regions are assumed to ne self-shielded in the CF reionization model). The mean free path can be used to calculate the redshift distribution $\de N_{\mathrm{LL}} / \de z$ of the Lyman-limit systems, which can then be compared with the corresponding observations \citep{2011ApJ...736...42R, 2013ApJ...765..137O, 2013ApJ...775...78F, 2010ApJ...718..392P, 2019MNRAS.482.1456C} to obtain constraints on $\lambda_{0}$. 

\item The hydrogen photoionization rate $(\Gamma_{\mathrm{PI}})$ can be obtained from the photon production rate $\dot{n}_{\mathrm{ph}}(z)$ and the mean free path $\lambda_{\mathrm{mfp}}$ using
 \begin{equation}
      \Gamma_{\mathrm{PI}}(z)=(1+z)^{3} \int^{\infty}_{\nu_H} \de \nu~ \lambda_{\mathrm{mfp}}(\nu,z)~\dot{n}_{\mathrm{ph}}(z)~\sigma_{H}(\nu),
 \end{equation}
 where $\sigma_{H}(\nu)$ is the hydrogen photoionization cross-section. Similar equations can be used for calculating the photoionization rate for HeII as well.
\end{enumerate}

The CF reionization has been applied in several different kinds of studies, e.g., constraining escape fraction of ionizing photons from galaxies \citep{2013MNRAS.428L...1M}, quantifying the contribution of quasars to hydrogen reionization by including helium reionization data in the analysis \citep{2018MNRAS.473.1416M}, constraining reionization in tilted flat and untilted non-flat dynamical dark energy inflation models \citep{2018MNRAS.479.4566M,2019MNRAS.487.5118M}, constraining the primordial magnetic fields \citep{2015MNRAS.451.1692P}.

\subsection{The global 21~cm signal from cosmic dawn}

The 21~cm global brightness temperature is calculated following the calculations of \citet{2006MNRAS.371..867F,2019MNRAS.487.3560C}. The sky averaged 21~cm brightness temperature can be calculated using \citep{2019MNRAS.487.3560C}
\begin{equation}
    \delta T_b(\nu) \approx 10.1~\mathrm{mK}~x_{\mathrm{HI}}(z) \left[1 - \frac{T_{\gamma}(z)}{T_S(z)}\right]~(1+z)^{1/2},
\end{equation}
where $T_{\gamma}$ is the background radiation temperature, $T_{S}$ is the neutral hydrogen spin temperature and $x_{\mathrm{HI}}$ is the neutral hydrogen fraction in the IGM. The spin temperature $T_{S}$ is computed using \citep{1958PIRE...46..240F} 
\begin{equation}
    T_S^{-1}=\frac{T_{\gamma}^{-1}+x_c T_{K}^{-1}+x_{\alpha}T_\alpha^{-1}}{1+x_c+x_{\alpha}},
    \label{eq:T_s}
\end{equation}
where $T_{K}$ is the kinetic temperature of the IGM, $T_{\alpha}$ is the Ly$\alpha$ color temperature and $x_{c}$ and $x_{\alpha}$ are respectively the collisional and Ly$\alpha$ coupling coefficients. As the optical depth of the Ly${\alpha}$ is very high in the epoch of cosmic dawn, $T_{\alpha}$ rapidly approaches $T_{K}$. Moreover, in the redshift range observed using EDGES where the 21~cm signal calculations are most relevant, the collisions turn out to be negligible $x_c \ll 1$ \citep{2006MNRAS.371..867F}. With these approximation, \eqn{eq:T_s} simplifies to 
\begin{equation}
    T_S^{-1}=\frac{T_{\gamma}^{-1} + x_{\alpha} T_K^{-1}}{1+x_{\alpha}}.
\end{equation}
In the CF reionization model, the gas kinetic temperature is determined mainly by two processes, namely, the adiabatic cooling and the photoheating from UV photons. During the cosmic dawn, however, the heating is dominated by X-rays arising from star formation. To account for this fact, we include the X-ray heating term in the temperature evolution equation. Since the X-ray heating tends to become sub-dominant compared to the UV heating once the reionization starts, we switch off the X-ray heating in the reionization epoch. This allows us to keep the code numerically efficient.

To model the X-ray heating, we assume that the X-ray emissivity is proportional to the star formation rate. This is motivated by the observational result that the star formation rate and the X-ray luminosity of galaxies observed in the local Universe are strongly correlated  \citep{2012MNRAS.419.2095M}. We assume that such a correlation holds at high-$z$ and write the X-ray energy injection rate per unit comoving volume as
\begin{equation}
\frac{\epsilon_{X}}{\mathrm{J}~\mathrm{s}^{-1} \mathrm{Mpc}^{-3}} = 3.4 \times 10^{33} \frac{\rho_{b} \left[f^{\mathrm{II}}_{*} f^{\mathrm{II}}_{Xh} \frac{\de f_{\mathrm{coll, II}}}{\de t} +  f^{\mathrm{III}}_{*} f^{\mathrm{III}}_{Xh} \frac{\de f_{\mathrm{coll, III}}}{\de t}\right]}{M_{\odot} \mathrm{Mpc}^{-3} \mathrm{yr}^{-1}},
\end{equation}
where $f_{Xh}=f_{X}~f_{h}$ (for both PopII and PopIII stars). The unknown normalization factor $f_{X}$ takes into account any difference between local observations and high redshift, and $f_{h}$ is the heating fraction of the total X-ray photons in the IGM (the other part helps in ionization).

The Ly$\alpha$ coupling coefficient $x_{\alpha}$ is determined by the background Ly$\alpha$ flux $J_{\alpha}$ through the relation
\begin{equation}
x_{\alpha}=\frac{1.81 \times 10^{11}}{ (1+z)} S_{\alpha} \frac{J_{\alpha}} {\mathrm{cm}^{-2} \mathrm{s}^{-1} \mathrm{Hz}^{-1} \mathrm{sr}^{-1}},
\end{equation}
where $S_{\alpha}$ is a factor of order unity. It accounts for the detailed atomic physics involved in the scattering process. For simplicity, we fix $S_{\alpha}=1$. The flux can be calculated from the photon production rate as
\begin{equation}
J_{\alpha}(z) =\frac{c}{4 \pi}(1+z)^3 \int^{z_{\mathrm{max}}}_{z} \de z'~\left[f^{\mathrm{II}}_{\alpha}~\dot{n}_{\nu'}^{\mathrm{II}}(z') + f^{\mathrm{III}}_{\alpha}~\dot{n}_{\nu'}^{\mathrm{III}}(z')\right] \left|\frac{\de t'}{\de z'}\right|,
\end{equation}
where $f_{\alpha}$ (defined for both the stellar populations) is a normalization parameter that takes into account any uncertainties in the Ly$\alpha$ flux that may arise from the unknown properties of the high redshift galaxies. For example, it is possible that some fraction of Ly$\alpha$ photons may not escape into IGM because of some unknown interaction within the ISM, in that case the uncertainty will get absorbed in this parameter. The upper limit $z_{\mathrm{max}}$ of the integral accounts for the fact that continuum ionizing would be absorbed in the IGM and not contribute the Ly$\alpha$ radiation. The limit is calculated using \citep{2020MNRAS.496.1445C} 
\begin{equation}
    1+z_{\mathrm{max}}=\frac{\nu_{H}}{\nu_{\alpha}} (1+z),
\end{equation}
where $\nu_{\alpha}$ is the Ly$\alpha$ frequency.\footnote{The Ly$\alpha$ radiation also leads to heating of the IGM \citep[see, e.g.,][]{2020MNRAS.492..634G, 2021MNRAS.503.4264M} which is not incorporated in our model. Since both the Ly$\alpha$ flux and the heating are driven by the collapse rate $\de f_{\mathrm{coll}} / \de t$, inclusion of this process should affect the constraints on the efficiency parameters related to the stars. However, we do not expect the constraints on the cosmological parameters to be affected in any significant manner.} Note that the photon production rate $\dot{n}_{\nu}$ is proportional to $f_*$, see \eqn{eq:ndot_II_III}, hence $J_{\alpha}$ is proportional to the combination $f_* f_{\alpha}$ of the two populations.

It is well-known that the amplitude of the EDGES absorption signal \citep{2018Natur.555...67B} cannot be explained by standard models of galaxy formation. There have been discussions that the cosmological signal implied by EDGES may be an artefact of different systematics \citep{2018Natur.564E..32H, 2019ApJ...874..153B, 2019ApJ...880...26S, 2020MNRAS.492...22S}. While the confirmation of the result will require more tests with different instruments, we, for the time being, assume that the cosmological 21~cm signal is indeed that implied by \citet{2018Natur.555...67B}. Various speculations have been made in order to explain the excess dip of the EDGES signal, e.g., some exotic dark matter models where the dark matter particles interact with the baryons \citep{2018PhRvD..98j3005B, 2018PhLB..785..159F, 2018PhRvL.121c1103P, 2018PhRvD..98b3013S} and an excess radio background driven by formation of early stars \citep{2018ApJ...858L..17F, 2018ApJ...868...63E, 2019MNRAS.483.1980M, 2019MNRAS.486.1763F, 2020MNRAS.493.1217M, 2020MNRAS.492.6086E}. We avoid exploring any non-standard dark matter scenarios in this work and consider the latter scenario of excess radio flux. Note that this assumption may have some observational support from the excess observed by the ARCADE-2 experiment \citep{2011ApJ...734....5F}, although the origin of the excess is somewhat debatable \citealt{2011ApJ...734....6S}. The radio background is modelled based on the work of \cite{2019MNRAS.487.3560C} and calculated using
\begin{equation}
\frac{\epsilon_{R}}{\mathrm{J}~\mathrm{s}^{-1}~\mathrm{Mpc}^{-3}} =10^{22} \frac{\rho_{b} \left[ f^{\mathrm{II}}_{*} f^{\mathrm{II}}_{R} \frac{\de f_{\mathrm{coll, II}}}{\de t} +  f^{\mathrm{III}}_{*} f^{\mathrm{III}}_{R} \frac{\de f_{\mathrm{coll, III}}}{\de t}\right]}{M_{\odot}~\mathrm{Mpc}^{-3}~\mathrm{yr}^{-1}},
\end{equation}
where, $f_{R}$ takes into account the uncertainty one needs to introduce in order to extrapolate the radio-SFR relation from local to high redshift Universe.

It is worth noting that there is still no self-consistent model of producing a strong radio background at high redshifts. There have been concerns whether the accelerating relativistic electrons that produce the excess background can be sustained as their cooling time is much shorter than the Hubble time at $z \sim 17$ \citep{2018MNRAS.481L...6S}. On the other hand, \citet{2020MNRAS.492.6086E} have pointed out that the continuous injection of particles leads to a different time-evolution of the radiation spectrum and hence the concerns may not be that severe.

\subsection{The CMB anisotropies}

Given the reionization model, we need to compute the CMB angular anisotropies, in particular, the angular power spectra $C_l$ for the temperature and E-mode polarization fluctuations. For this purpose, we use the publicly available python-wrapped CAMB \cite{Lewis:2013hha}\footnote{\href{https://camb.readthedocs.io/en/latest/}{https://camb.readthedocs.io/en/latest/}}, which is an efficient code for computing the CMB anisotropies. The default reionization history used in the CAMB input is the simple tanh parametrization \cite{2008PhRvD..78b3002L}, which is not sufficient for our work. As was done by \citet{2012MNRAS.419.1480M}, we modify CAMB to incorporate the reionization history implied by our reionization model. The resulting output can be compared with the CMB data sets in a straightforward manner.

\begin{figure*}
  \includegraphics[width=\textwidth]{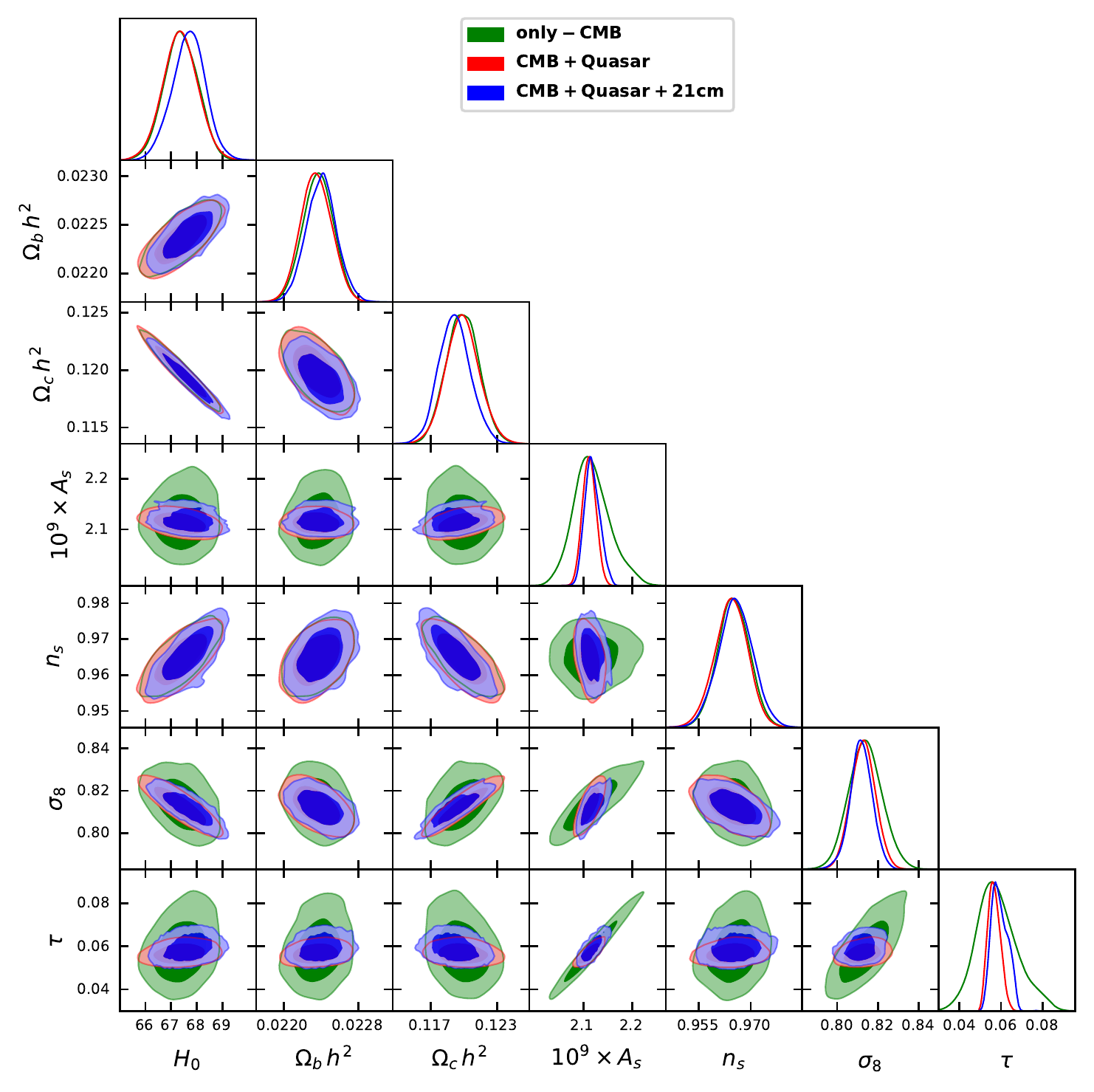}
    \caption{The marginalized posterior distributions of cosmological parameters obtained from \textbf{only-CMB} (green), \textbf{CMB+Quasar} (red) and \textbf{CMB+Quasar+21cm} (blue) analyses. We show the $68 \%$ and $95\%$ confidence contours in the two-dimensional plots. Note that $\sigma_8$ and $\tau$ are derived parameters while the others are free parameters in our model. By comparing the \textbf{only-CMB} and \textbf{CMB+Quasar} results, it is clear that the inclusion of the quasar absorption data in the analysis leads to much tighter constraints on $\tau$ and consequently on $A_s$ and $\sigma_8$.}
    \label{fig:all}
\end{figure*}

\subsection{The \texttt{CosmoReionMC} Package}

Finally, we combine the various components of the model described in the preceding subsections and combine with a module for carrying out the likelihood analysis. This allows us to do parameter estimation in an efficient manner. The resulting package is named as \texttt{CosmoReionMC}. As discussed later, this package has many potential applications, e.g., put constraints on non standard cosmology, constraining the mass of dark matter particles, non standard dark energy, non-flat cosmological models and many more. We plan to make this package public in the near future.

The likelihood analysis to constrain the parameters is based on MCMC. For this purpose, we develop a module which is inspired from the publicly available \texttt{CosmoHammer} \citep{2013A&C.....2...27A} for estimating cosmological parameters. The MCMC routine is driven by the publicly available \texttt{emcee} code \citep{2013PASP..125..306F}.

In order to connect with the latest python-wrapped CAMB and \texttt{emcee}, the reionization and 21~cm codes have now been written completely in python.\footnote{As is well known, MCMC analysis requires executing the core code millions of times, hence it needs to be optimized in time. Here, we use \emph{Numba} (\href{http://numba.pydata.org.}{http://numba.pydata.org.}) to make the code much faster so that the execution time becomes somewhat comparable to that of Fortran.} We have ensured that the code is optimized in terms of computing efficiency. For example, a typical MCMC run takes about 24 hours using 24 cores.

So to summarize, our parameter estimation package \texttt{CosmoReionMC} is an efficient combination of (i) a semi-analytical reionization and cosmic dawn model \citep{2005MNRAS.361..577C,2019MNRAS.487.3560C}, (ii) CMB anisotropy code CAMB \citep{Lewis:2013hha} modified to account for arbitrary reionization histories and (iii) the \texttt{emcee} code \citep{2013PASP..125..306F} for the statistical MCMC analysis.

\section{Parameter constraints}
\label{section:reionization_CMB}

We now present the results on the parameter constraints obtained using \texttt{CosmoReionMC}.

\begin{table*}

\renewcommand{\arraystretch}{2.0}

\begin{tabular}{|c|c|c|c|c|c|c|c|}
\hline
Parameter & Prior & \multicolumn{2}{c|}{only-CMB} & \multicolumn{2}{c|}{CMB+Quasar} & %
    \multicolumn{2}{c|}{CMB+Quasar+21cm} \\

\cline{3-8}
& & best-fit & mean [95$\%$ C.L.] & best-fit & mean [95$\%$ C.L.] & best-fit & mean [95$\%$ C.L.]  \\
\hline
 $H_{0}$ &[65, 80] & 67.21 & 67.39 [66.10, 68.65] & 67.24 & 67.36 [66.04, 68.70] & 67.51 & 67.68 [66.37, 68.94] \\

 $\Omega_b h^2$ & [0.005, 0.05] & 0.02234 & 0.02237 [0.02205,  0.02268] & 0.02234 & 0.02234 [0.02203,  0.02267] & 0.02236 & 0.02240 [0.02208,  0.02272 ] \\

 $\Omega_c h^2$ & [0.01, 0.2] & 0.1202 & 0.1198 [0.1170,  0.1228 ] & 0.1201 & 0.1198 [0.1169,  0.1229 ] & 0.1198 & 0.1191 [0.1164,  0.1221 ] \\

$10^{9}A_{s}$ &[1.0, 3.0]  &  2.092 & 2.117 [2.048, 2.200] & 2.102 & 2.112 [2.085,  2.139] & 2.105 & 2.118 [2.091,  2.151] \\

$n_{s}$ & [0.8, 1.2] &  0.9627 & 0.9650 [0.9559,  0.9739] & 0.9625 & 0.9643 [0.9547,  0.9734] & 0.9640 & 0.9656 [0.9557,  0.9754] \\

$f^{\mathrm{II}}_{\mathrm{esc}}$  & [0.0, 1.0] & 0.1372 & 0.1568 [0.1168, 0.3662] &  0.1681 & 0.1951 [0.1498, 0.2471] & 0.1723 & 0.2096 [0.1514 , 0.2850] \\

$\lambda_{0}$ & [1.0, 10.0] & 1.53 & 3.96 [1.0, 9.99] & 4.38 & 3.97 [2.58,  5.47] & 3.97 & 3.63 [2.06, 5.20] \\

$f^{\mathrm{II}}_{Xh}$ & [0.0, 100.0] & $-$ & $-$ & $-$ & $-$ & 4.56 & 4.30 [1.93,  7.28] \\

$f^{\mathrm{III}}_{*, \mathrm{esc}}$ & [0.0, 1.0] & $-$ & $-$ & $-$ & $-$ & 0.0024 & 0.0030 [0.0008, 0.0055] \\

$f^{\mathrm{III}}_{*, \alpha}$ & [0.0, 50.0] & $-$ & $-$ & $-$ & $-$ & 0.534 & 0.700 [0.157, 2.91] \\

$f^{\mathrm{III}}_{*, R}$ & [$10^{3}$, $10^{7}$] & $-$ & $-$ & $-$ & $-$ & 55319.2 & 53519.8 [24619.8, 89774.5] \\

$z_{\mathrm{trans}}$ & [9.0, 25.0] & $-$ & $-$ & $-$ & $-$ & 16.35 & 16.27 [15.89, 16.68] \\

\hline
$\sigma_{8}$ & $-$ & 0.809 & 0.814 [0.798,  0.830] & 0.811 & 0.812 [0.801,  0.824] & 0.812 & 0.812 [0.802,  0.823] \\

$\tau$ & $-$ & 0.0512 & 0.0581 [0.0402,  0.0802] & 0.0533 & 0.0557 [0.0512,  0.0624] & 0.0591 & 0.0618 [0.0529,  0.0670] \\
\hline
\multicolumn{2}{|c|}{$\mathcal{L}_{\mathrm{pl}}$}  & 693.50 & & 701.50 & & 702.40 & \\
\hline
\multicolumn{2}{|c|}{$\mathcal{L}_{\mathrm{Re}}$}  & $-$ & & 3.05 & & 3.67 & \\
\hline
\multicolumn{2}{|c|}{$\mathcal{L}_{21}$}  & $-$ & & $-$ & & 201.91 & \\
\hline
\end{tabular}
\caption{Constraints on the model parameters for the three different analyses of the paper. For each parameter, we show the prior along with the best-fit, mean and $95\%$ confidence limits. The parameters $\sigma_8$ and $\tau$ are derived. The last three rows show the best-fit log-likelihood values for the different data sets.
}
\label{tab:table_1}
\end{table*}

\subsection{Constraints using only CMB data}
\label{section:only_CMB}

We first apply our code to obtain constraints on the cosmological parameters using only CMB data. For this purpose, we compare the CMB angular power spectrum predicted by our model with the Planck 2018 observations. The free parameters for this analysis (referred to as \textbf{only-CMB} hereafter) are
\begin{equation}
    \Theta =\{H_{0}, \Omega_{b}h^{2}, \Omega_ch^{2}, A_{s}, n_s, f^{\mathrm{II}}_{\mathrm{esc}}, \lambda_{0}  \},
\end{equation}
where the first five parameters are the usual cosmological parameters in the flat $\Lambda$CDM model and the last two are the free parameters of our reionization model. We ignore the contribution of PopIII stars in this part of the analysis as it has been shown \citep {2020MNRAS.tmp.2712Q, 2015MNRAS.454L..76M, 2011MNRAS.413.1569M} that they are not required for matching the reionization-related observations. To confirm this, we have also run a model including PopIII stars and found that the constraints are consistent with no contribution from PopIII stars. We also found that the constraints on the other parameters are not affected even if we ignore the PopIII contribution.

The main input to the MCMC analysis is the negative of the log-likelihood denoted as $\mathcal{L}$ (so that the likelihood is $L \propto \exp \left(- \mathcal{L} \right)$). For the \textbf{only-CMB} analysis, this is given by
\begin{equation}
    \mathcal{L}=\mathcal{L}_{\mathrm{Pl}},
\end{equation}
where $\mathcal{L}_{\mathrm{Pl}}$ is the log-likelihood function corresponding to the Planck 2018 observations. For $l \geq 30$, the joint likelihood for TT, TE and EE fluctuations are computed using the ``Plik lite'' method where the foreground and other nuisance parameters are marginalized over appropriate prior ranges. For small-$l$ values, we use the likelihoods obtained by the "Commander" approach for TT \citep{2014A&A...571A..15P} and by the "SimAll" algorithm for EE and BB.\footnote{The details of these likelihoods can be found on the wiki page https://wiki.cosmos.esa.int/planck-legacy-archive/index.php/CMB\_spectrum\_\%26\_Likelihood\_Code. In particular, we use\\
\texttt{hi\_l/plik\_lite/plik\_lite\_v22\_TTTEEE.clik} for the high-$l$ likelihood,\\ 
\texttt{low\_l/commander/commander\_dx12\_v3\_2\_29.clik} for the low-$l$ TT,\\
\texttt{low\_l/simall/simall\_100$\times$143\_offlike5\_EE\_Aplanck\_B.clik} for the low-$l$ EE and\\
\texttt{low\_l/simall/simall\_100$\times$143\_offlike5\_BB\_Aplanck\_B.clik} for the low-$l$ BB fluctuations.}

The other input to the MCMC analysis is the prior on the free parameters. We assume a broad flat prior for all the seven free parameters, the values are mentioned in ``Prior'' column of Table~\ref{tab:table_1}. For exploring the parameter space with MCMC chains, we use 32 walkers taking $10^{6}$ steps. The first $30\%$ steps are removed from the analysis as ``burn-in'' and the posterior distributions are computed based on the remaining $70\%$ steps. An auto-correlation analysis has been carried out for ensuring the convergence of the chains. We have calculated the integrated autocorrelation time $\tau_f$ for each parameter-chain (different steps of walkers corresponding to a particular free parameter). We use \texttt{autocorr.integrated\_time} function of the \texttt{emcee} package for this purpose. A parameter-chain is said to be converged if the length of the chain is more than $100 \times \tau_{f}$ following a conservative approach discussed in \cite{2013PASP..125..306F}.  

The posterior distributions of the model parameters are shown in \fig{fig:all} (green contours). The constraints on the parameters, i.e., the best-fit values along with the mean and 95\% C.L are shown in Table~\ref{tab:table_1}. We also show the log-likelihood for the best-fit model in the same table. Among the various correlations between the cosmological parameters, one which will play an important role in the subsequent discussions is that between $A_s$ and $\tau$ (or equivalently between $\sigma_8$ and $\tau$). The reason for this tight correlation arises because the amplitudes of the CMB angular power spectra $C_l$ are proportional to the combination $A_s~{\rm e}^{-2 \tau}$. Since the reionization history is closely coupled to the value of $\tau$, one anticipates that the inclusion of reionization-related data from quasar absorption spectra would affect the constraints on $\tau$ and hence $A_s$ and $\sigma_8$.

This \textbf{only-CMB} analysis will act as our baseline against which we will compare the rest of the analyses. In particular, these constraints will allow us to understand the effect of including reionization and 21~cm data in the next sections.

\subsection{Joint constraints using CMB and quasar absorption data}

\begin{figure*}
    \centering
    \includegraphics[width=\textwidth]{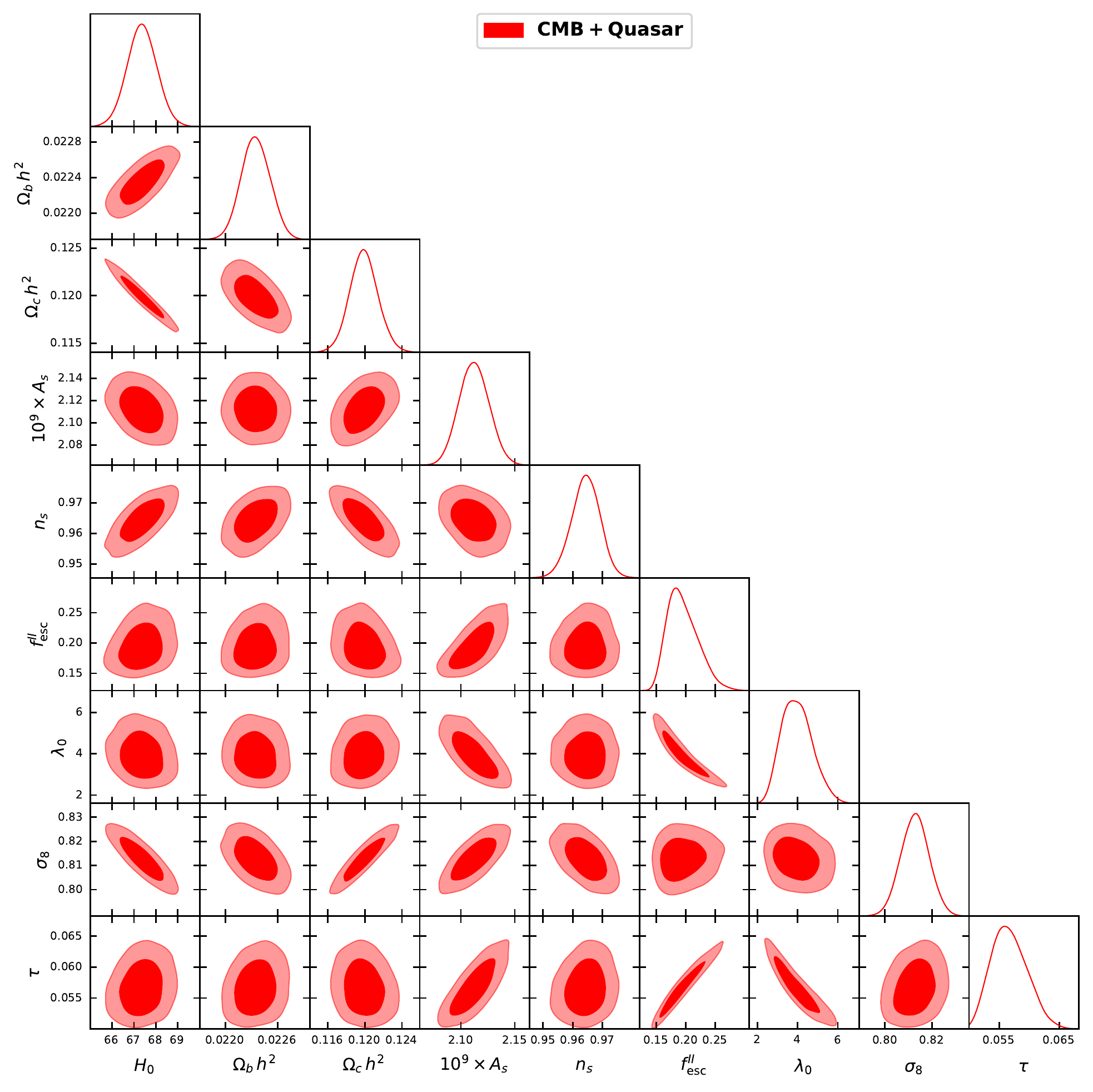}
    \caption{The marginalized posterior distribution of seven free parameters and two derived parameters ($\sigma_8$ and $\tau$) obtained for the \textbf{CMB+Quasar} case. We show the $68\%$ and $95\%$ confidence contours in the two-dimensional plots. Note that the posteriors for the cosmological parameters are identical to those for the \textbf{CMB+Quasar} case shown in \fig{fig:all}.}
    \label{fig:getdist_reionization}
\end{figure*}

\begin{figure*}
    \centering
    \includegraphics[width=\textwidth]{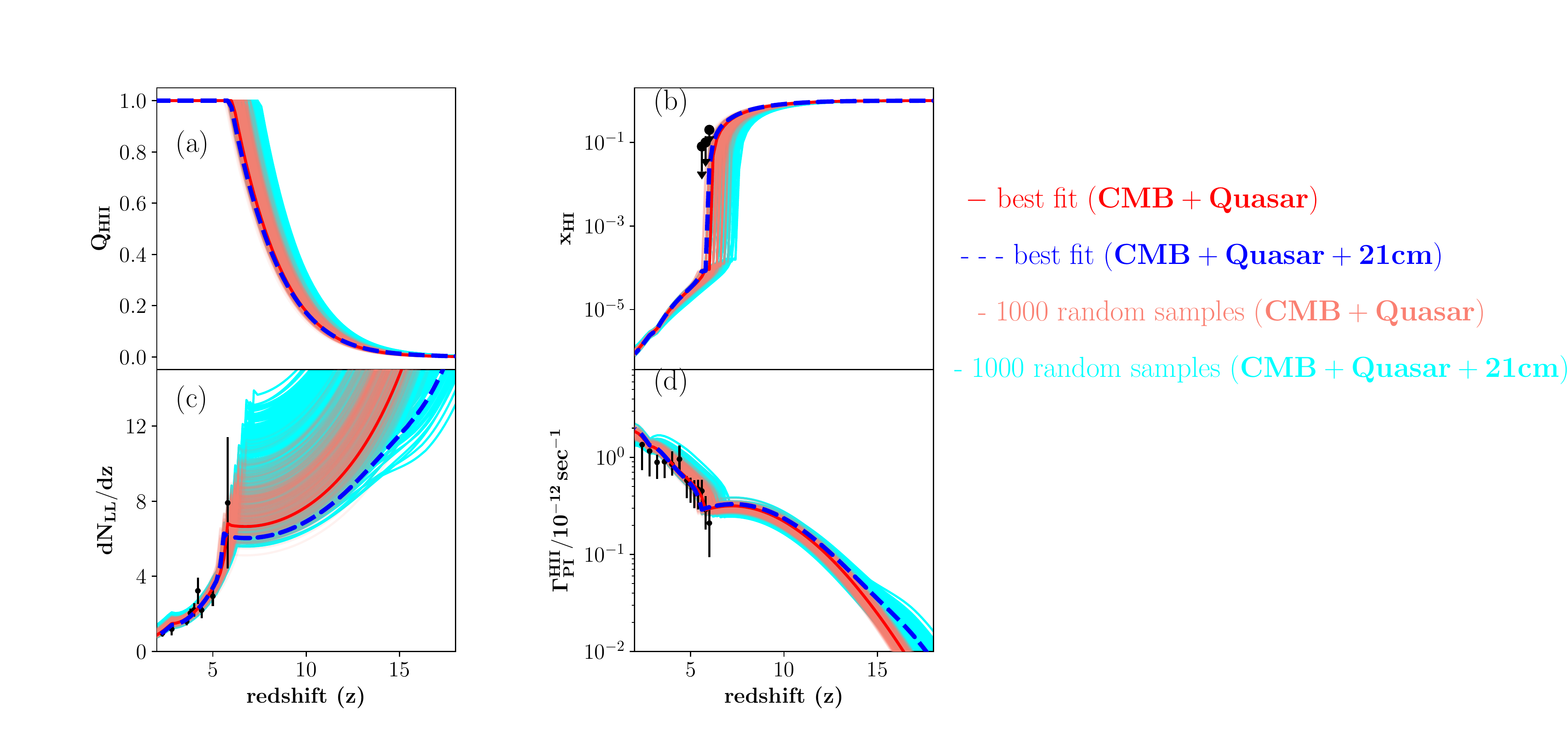}
    \caption{Redshift evolution of different quantities related to reionization for both \textbf{CMB+Quasar} and \textbf{CMB+Quasar+21cm} cases. In each panel, the light-red region show 1000 random samples obtained from the MCMC chains for the \textbf{CMB+Quasar} analysis whereas the red lines show the corresponding best-fit model. The cyan shaded region shows 1000 random samples for the \textbf{CMB+Quasar+21cm} analysis and the blue dashed curves show the coresponding best-fit curves. The different panels are: (a) The volume filling factor of HII region ($Q_{\mathrm{HII}}$). (b) The neutral hydrogen fraction ($x_{\mathrm{HI}}$). The black points with error-bars show the data points obtained from dark pixels fraction of Ly$\alpha$ absorption spectra \citep{2015MNRAS.447..499M}. (c) The redshift distribution of Lyman-limit system $\de N_{\mathrm{LL}} / \de z$. The black points with error-bars show the observational data points \citep{2010ApJ...721.1448S,2019MNRAS.482.1456C}. (d) The hydrogen photoionization rate $(\Gamma_{\mathrm{PI}})$. The black points with error bars are the observational data points \citep{2011MNRAS.412.2543C,2013MNRAS.436.1023B,2018MNRAS.473..560D}.  }
    \label{fig:redshift_reionization}
\end{figure*}

Next, we include the reionization related observations from quasar absorption spectra in our analysis along with the Planck 2018 observations. The additional data sets related to reionization used for this analysis are described below:

\begin{enumerate}
    
    \item Measurements of the photoionization rate $\Gamma_{\mathrm{PI}}$ obtained from a combined analysis of quasar absorption spectra and hydrodynamical simulations  \citep{2013MNRAS.436.1023B, 2018MNRAS.473..560D,2011MNRAS.412.2543C}.
        
    \item The redshift distribution of Lyman-limit system $\de N_{\mathrm{LL}} / \de z$ \citep{2011ApJ...736...42R, 2013ApJ...765..137O, 2013ApJ...775...78F,  2010ApJ...718..392P, 2019MNRAS.482.1456C, 2010ApJ...721.1448S}. To calculate $\de N_{\mathrm{LL}} / \de z$ from our model, we need to assume a functional form for the frequency distribution of the HI column density $(N_{\mathrm{HI}})$ and then have to integrate this function over all relevant $N_{\mathrm{HI}}$ values. The column density distribution has been taken from the latest compilation by \cite{2019MNRAS.482.1456C}. The lower limit of this integration depends on how one defines the Lyman-limit systems. For the $2<z<5$, the observational data points are taken from \cite{2019MNRAS.482.1456C} where they have defined the Lyman-limit systems with $N_{\mathrm{HI}} \ge 10^{17.5}$~cm$^{-2}$. Hence in this redshift range, we take the lower limit of the integration to be $10^{17.5}$~cm$^{-2}$. On the other hand, the observational point at $z \sim 6$ has been taken from \cite{2010ApJ...721.1448S} where all systems with $N_{\mathrm{HI}} \ge 10^{17.2}$~cm$^{-2}$ are considered as Lyman-limit. Hence we have taken the lower limit of the integration to be $10^{17.2}$~cm$^{-2}$ for $z>5$.  

    \item The model independent upper limit on the neutral hydrogen fractions, measured from the dark fractions in quasar spectra \citep{2015MNRAS.447..499M}.

    \item We also put a prior that reionization needs to be completed ($Q_{\mathrm{HII}} = 1$) at $z \geq 5.3$. This redshift constraint is motivated by studies of the large scale fluctuations of the effective Ly$\alpha$ optical depth from high redshift quasar spectra \citep{2015MNRAS.447.3402B,2018MNRAS.479.1055B,2017ApJ...840...24E,2018ApJ...864...53E}. For example, using radiative transfer simulation, \citep{2019MNRAS.485L..24K} predicts the completion of reionization to take place at redshift $z \approx 5.3$. Similarly, \cite{2020arXiv200308958C} used a semi numerical approach to study the same and their model also predicted similar redshift for the completion of reionization.
\end{enumerate}

The analysis procedure (referred as \textbf{CMB+Quasar} hereafter) is very similar to the \textbf{only-CMB} analysis. However, as we include the reionization observations, the log-likelihood is modified to
\begin{equation}
    \mathcal{L}=\mathcal{L}_{\mathrm{Pl}} + \mathcal{L}_{\mathrm{Re}},
\end{equation}
where
\begin{equation}
\mathcal{L}_{Re}=\frac{1}{2}\sum^{N_{\mathrm{obs}}}_{\alpha=1}\left[\frac{\zeta^{\mathrm{obs}}_{\alpha}-\zeta^{\mathrm{th}}_{\alpha}}{\sigma_{\alpha}}\right]^2.
\end{equation}
Here $\zeta^{\mathrm{obs}}_{\alpha}$ represents the set of $N_{\mathrm{obs}}$ observational data related to photoionization rates and the distribution of the Lyman-Limit system whereas $\zeta^{\mathrm{th}}_{\alpha}$ represents the values from the theoretical model. The $\sigma_{\alpha}$ are the observational error bars.

The posterior distribution and the constraints on different parameters are shown in \fig{fig:getdist_reionization}. The mean and $95 \%$ confidence interval of the different parameters used in the analysis can be found in Table~\ref{tab:table_1}. The correlation between the cosmological parameters are similar to what is expected from any standard analysis of the CMB data. Some of the other interesting points of our analysis are as follows:
\begin{enumerate}
    \item There is a strong positive correlation between $\tau$ and $f^{\mathrm{II}}_{\mathrm{esc}}$. This is obvious because as we increase the value of $f^{\mathrm{II}}_{\mathrm{esc}}$ the reionization will take place earlier and as a result $\tau$ will increase.
    
    \item The free parameter $\lambda_{0}$ shows a strong anti-correlation with $\tau$ and $f^{\mathrm{II}}_{\mathrm{esc}}$. This is because while calculating $\Gamma_{\mathrm{PI}}$ in the CF model, the product $\lambda_{0}~f^{\mathrm{II}}_{\mathrm{esc}}$ appears as a combination. This implies that, for a fixed value of $\Gamma_{\mathrm{PI}}$, if $f^{\mathrm{II}}_{\mathrm{esc}}$ is increased, $\lambda_0$ has to decrease so as to keep their product fixed. As we have already discussed that $f^{\mathrm{II}}_{\mathrm{esc}}$ is correlated with $\tau$, this immediately implies an anti-correlation between $\tau$ and $\lambda_0$.
\end{enumerate}

The constraints on the cosmological parameters are also shown in \fig{fig:all}, which allows a straightforward comparison with the previous \textbf{only-CMB} analysis. Similarly, one can also compare the obtained parameter limits for the two analyses from the respective columns in Table~\ref{tab:table_1}. Though the constraints on $H_{0}, \Omega_{b}h^{2}, \Omega_ch^{2}, n_s$ for the \textbf{CMB+Quasar} are not significantly different from the \textbf{only-CMB} analysis, the limits on $\tau$, $A_s$ and $\sigma_8$ are different. Firstly, we note that the constraint on $\tau$ is substantially tighter in \textbf{CMB+Quasar} analysis compared to that of \textbf{only-CMB}. This is a direct effect of including the quasar absorption observations in \textbf{CMB+Quasar} analysis. Lower values of $\tau$, which imply considerably delayed reionization, are disallowed by the data points from dark pixels and also the prior that reionization needs to be completed before $z = 5.3$. On the other hand, higher values of $\tau$ require higher values of $f_{\mathrm{esc}}^{\mathrm{II}}$ and hence tend to produce a $\Gamma_{\mathrm{PI}}$ higher than allowed by the data. These two effects lead to a much restricted range in $\tau$. Now, we have already discussed in the previous section that $A_s$ (or $\sigma_8$) and $\tau$ are correlated because of the dependence of $C_l$ on them, hence a narrower posterior distribution in $\tau$ will induce a similar change in $A_s$ and $\sigma_8$. For example, the 95\% C.L. for $\sigma_8$ is $\approx 0.023$ for the \textbf{CMB+Quasar} case, while it is $\approx 0.032$ for the \textbf{only-CMB} case.

\fig{fig:redshift_reionization} shows the redshift evolution of different quantities related to our reionization model: the ionized hydrogen fraction $Q_{\mathrm{HII}}$ (panel a), the neutral fraction $x_{\mathrm{HI}}$ (panel b), the redshift distribution of Lyman-limit systems $\de N_{\mathrm{LL}} / \de z$ (panel c) and the hydrogen photoionization rate $\Gamma_{\mathrm{PI}}$ (panel d). In all the panels, the red thick curves correspond to the best-fit model for the \textbf{CMB+Quasar} analysis whereas the light red region corresponds to 1000 random samples taken from the MCMC chains of this analysis. The black points are the observational data points. Our best-fit model agrees with the observational data points quite well. From panels (a) and (b), we see that the reionization starts around $z \sim 12$ and is complete by redshift $z \sim 5.5$.

We also mention the best-fit log-likelihood values corresponding to different analyses in Table~\ref{tab:table_1}. We find that the value of the best-fit $\mathcal{L}_{Pl}$ (the Planck CMB likelihood) for the \textbf{CMB+Quasar}  analysis is slightly higher than that of the \textbf{only-CMB}. This is because the best-fit model of the \textbf{only-CMB} analysis has a lower $\tau$ than what is preferred by the quasar absorption data sets; in fact, the reionization ends later than what is implied by the dark pixel data. This too shows how the inclusion of quasar absorption spectra affects the conclusions on allowed models.

\section{Joint analysis using CMB, quasar absorption and 21~cm data}
\label{section:CMB_reionization_21}

\begin{figure*}
  \includegraphics[width=\textwidth]{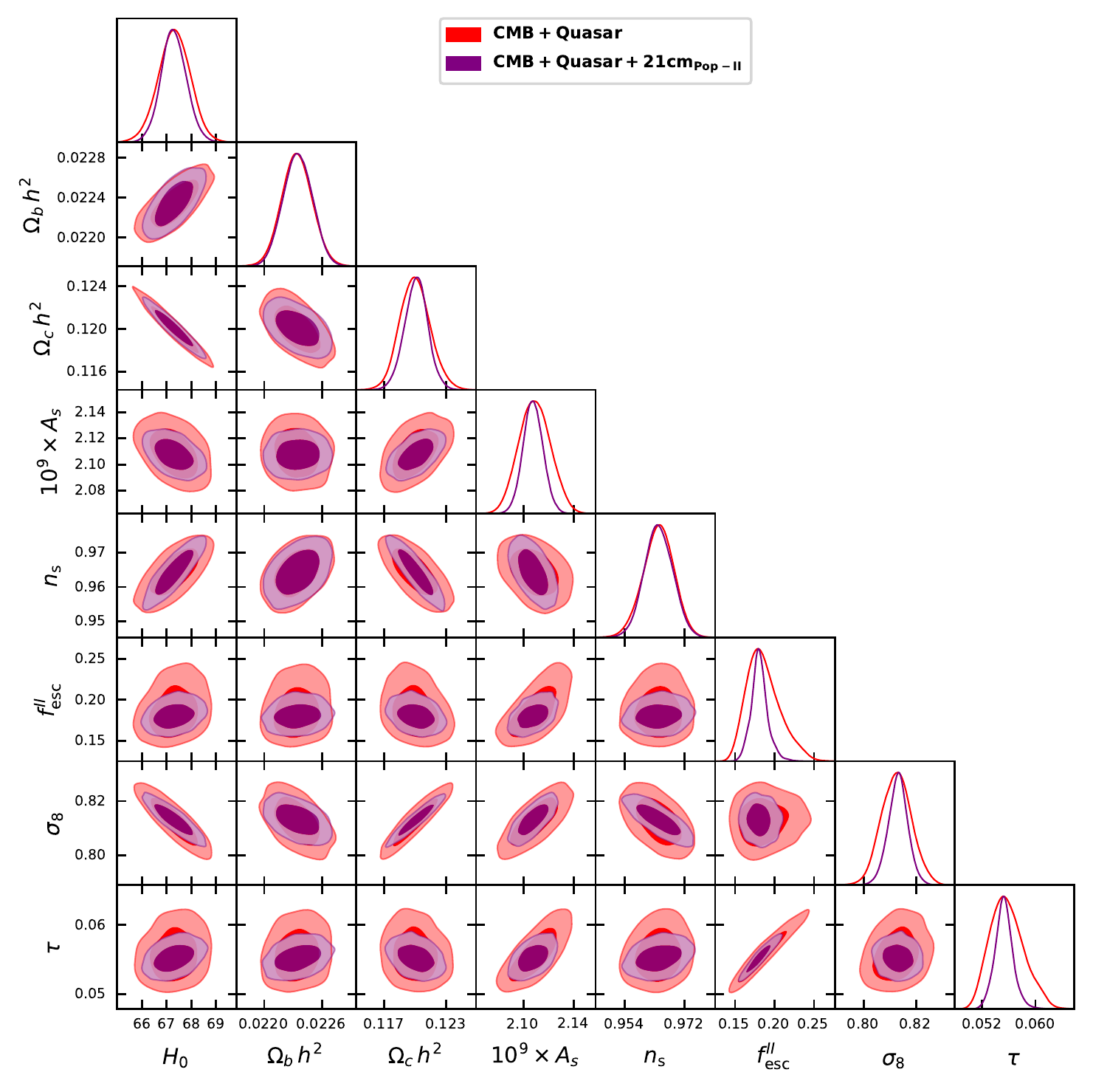}
    \caption{The marginalized posterior distributions of cosmological parameters obtained from  \textbf{CMB+Quasar} (red) and \textbf{CMB+Quasar+21cm$_{\rm \mathbf{Pop-II}}$} (purple) analyses. We show the $68 \%$ and $95\%$ confidence contours in the two-dimensional plots. Note that $\sigma_8$ and $\tau$ are derived parameters while the others are free parameters in our model.}
    \label{fig:all_pop2}
\end{figure*}

\begin{figure*}
    \centering
    \includegraphics[width=\textwidth]{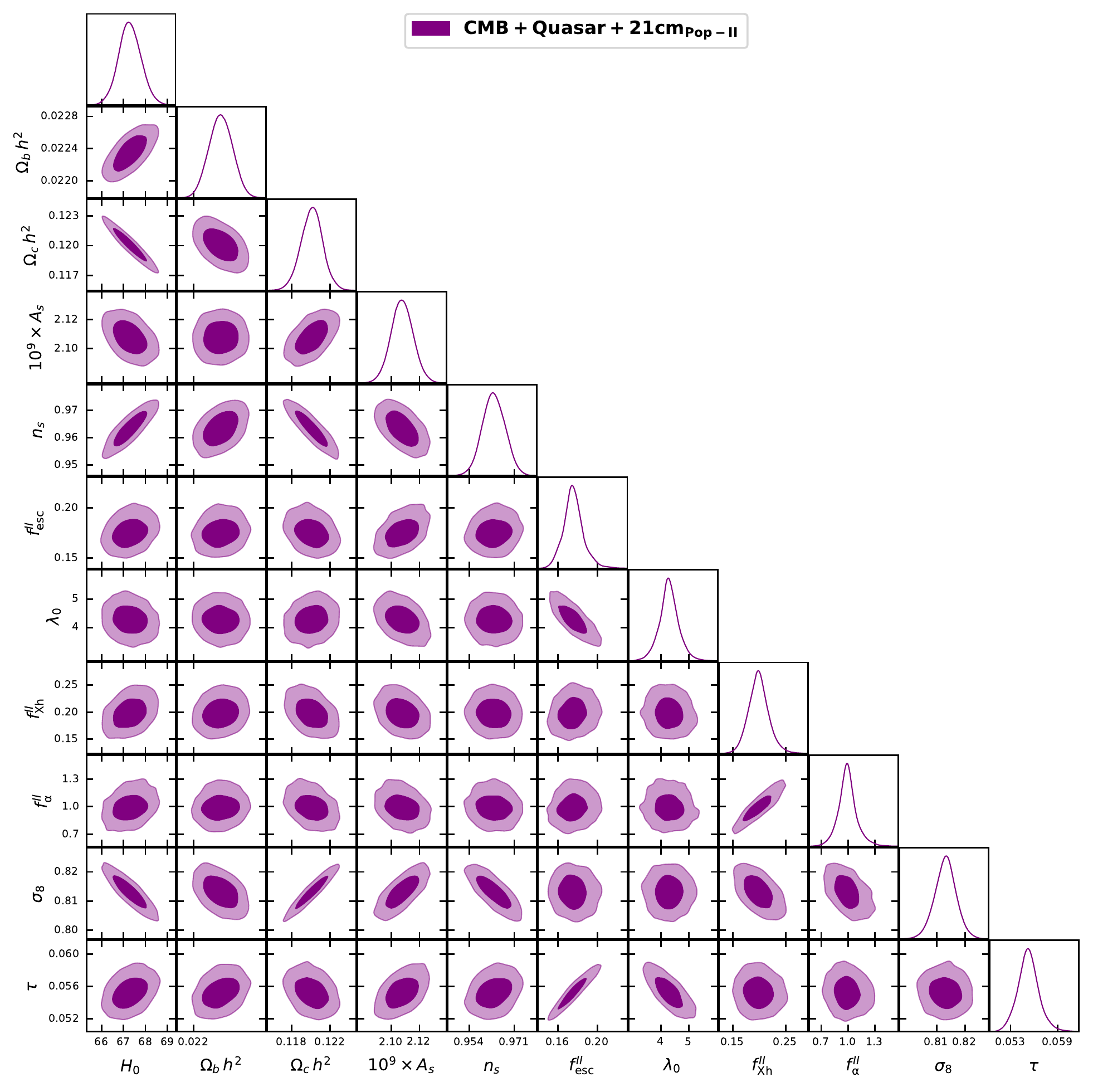}
    \caption{The marginalized posterior distribution of nine free parameters and two derived parameters ($\sigma_8$ and $\tau$) obtained for the \textbf{CMB+Quasar+21cm$_{\rm \mathbf{Pop-II}}$} case. We show the $68\%$ and $95\%$ confidence contours in the two-dimensional plots. Note that the posteriors for the cosmological parameters are identical to those for the \textbf{CMB+Quasar+21cm$_{\rm \mathbf{Pop-II}}$} case shown in \fig{fig:all_pop2}. }
    \label{fig:triangle_21_pop2}
\end{figure*}

\begin{figure}
    \centering
    \includegraphics[width=0.5\textwidth]{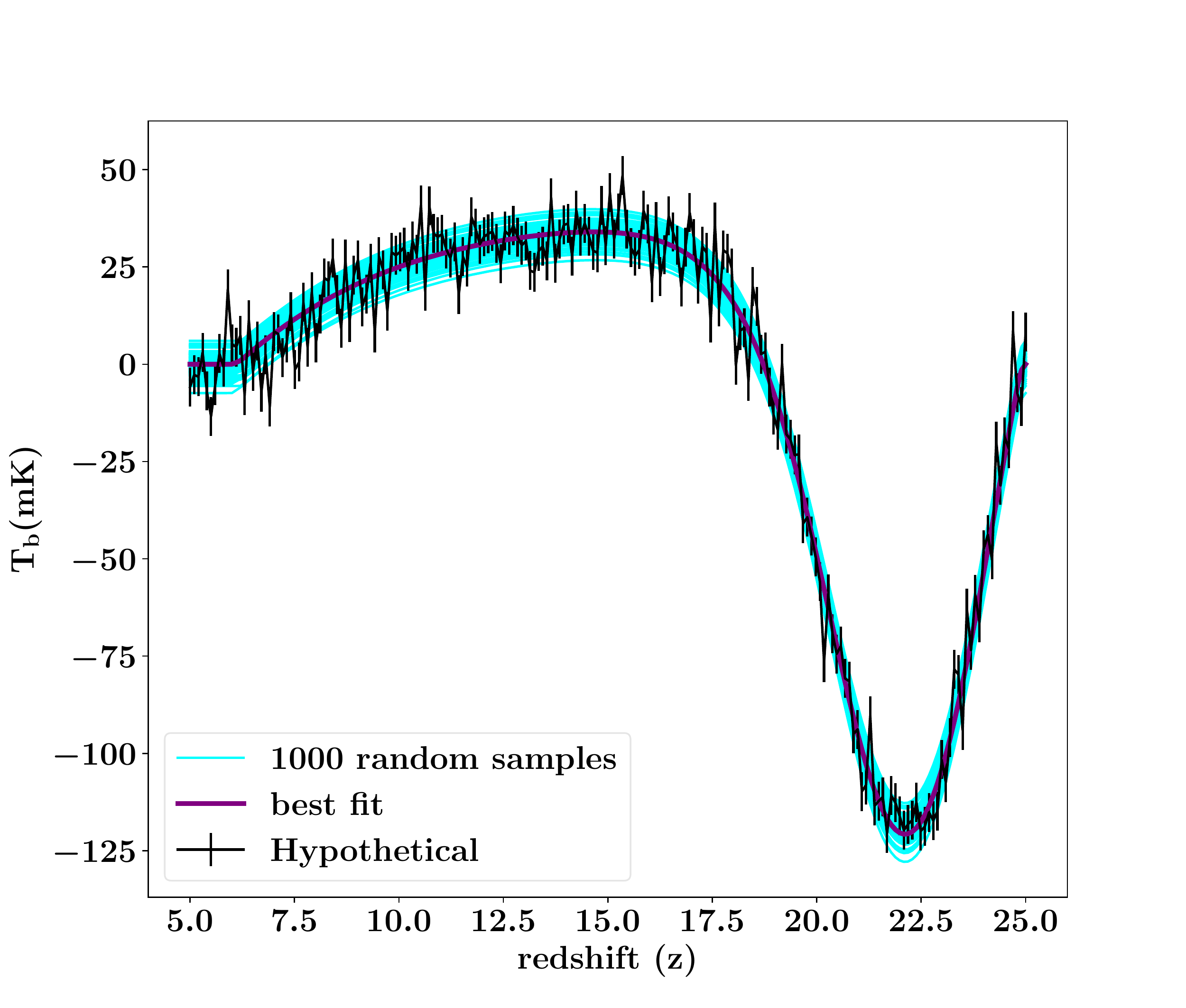}
    \caption{Comparison between the mock observations and the 21~cm signal recovered from our MCMC analysis in the \textbf{CMB+Quasar+21cm$_{\mathbf{Pop-II}}$} case. The purple curve shows the best-fit model whereas the cyan curves show 1000 random samples from the MCMC chains. The black line with the error bars represents the mock data points. Our recovered signal matches quite well with the mock data.}
    \label{fig:21cm_pop2}
\end{figure}

Next, for the first time, we include the 21~cm global signal data along with the CMB and reionization observations. Here we will discuss two scenarios, $(i)$ we use a hypothetical 21~cm signal as a mock observation and $(ii)$ we use the EDGES signal as the observations. The motivation behind doing the first analysis is that the EDGES signal requires including additional physics which leads to not so tight constraints on the cosmological parameters. The analysis with a hypothetical signal which is sensitive to reionization and does not require any additional physics allows us to demonstrate that using 21~cm signal could, in principle, be quite helpful in putting tighter constraints on cosmological parameters compared to \textbf{CMB+Quasar} analysis.

\subsection{CMB+Quasar+Hypothetical 21~cm signal}

In this analysis, we include a mock data for the global 21~cm signal in addition to the CMB and Quasar data. The basic procedure for this scenario (\textbf{CMB+Quasar+21cm$_{\rm Pop-II}$} hereafter) is similar to that of \textbf{only-CMB} and \textbf{CMB+Quasar}. The construction of the hypothetical signal will depend on the source model assumed. For the sake of simplicity, while constructing the hypothetical signal, we assume contribution only from the usual PopII stars and ignore any contribution from PopIII. This allows efficient computation of the parameter posterior distributions and also easier interpretation of the results.

The inclusion of the 21~cm signal in our analysis requires computation of the same from the model which leads to an increase in the number of free parameters. As we are taking contributions only from PopII stars, the additional free parameters would be
\begin{itemize}
    \item $f_{\alpha}^{\mathrm{II}}$: the normalization factor for the Ly$\alpha$ flux.
    \item $f_{Xh}^{\mathrm{II}}$: the heating efficiency of X-rays.
\end{itemize}

Thus, the nine free parameters for the joint analysis including the 21~cm signal are
\begin{equation*}
    \Theta =\{ H_{0}, \Omega_{b}h^{2}, \Omega_ch^{2}, A_{s}, n_s, f^{\mathrm{II}}_{\mathrm{esc}}, \lambda_{0}, f^{\mathrm{II}}_{Xh}, f^{\mathrm{II}}_{\alpha} \}
\end{equation*}
In the presence of the mock 21~cm observations, the log-likelihood becomes
\begin{equation}
     \mathcal{L}=\mathcal{L}_{\mathrm{Pl}} + \mathcal{L}_{\mathrm{Re}} + \mathcal{L}_{21, \rm PopII},
\end{equation}
where $\mathcal{L}_{21, \rm Pop-II}$, the loglikelihood corresponding to mock observational data is given by
\begin{equation}
    \mathcal{L}_{21, \rm Pop-II}= \sum_i \left[\frac{\delta T_b^{\mathrm{obs, PopII}}(\nu_i) - \delta T_b^{\mathrm{th}}(\nu_i)}{\sigma_i}\right]^2,
\end{equation}
where $\delta T_b^{\mathrm{obs, PopII}}(\nu_i)$ and $\delta T_b^{\mathrm{th}}(\nu_i)$ are the 21~cm differential brightness temperatures from the mock observational data and theoretical model, respectively and $\sigma_i$ are the corresponding observational errors. The sum is over all frequency channels $\nu_i$ of the mock data. We consider the frequency range 55-235 MHz, bin the signal into frequency channels of width 0.5 MHz. The mock data is generated using an assumed theoretical model (discussed in the next paragraph). In each bin, we add a Gaussian noise of zero mean and standard deviation $\sigma_i = 10$mK to the theoretical signal. This assumed value of the standard deviation is moderately better than that found in the EDGES analysis which can be achieved by a longer integration. This lower value of $\sigma_i$ is somewhat necessary to (i) detect the 21~cm emission signal during reionization (which is typically $\sim 30$~mK) and (ii) to put tighter constraints on cosmological parameters (compared to \textbf{CMB+Quasar}). With $\sigma_i \simeq 25$ mK (similar to EDGES), we find that varying the value of $\sigma_8$ or other cosmological parameters within the Planck allowed limit does not have any significant effect in the 21~cm signal.

As the main purpose of this analysis is to check whether the inclusion of 21~cm signal puts any tighter constraints on the cosmological parameters compared to the \textbf{CMB+Quasar}, we use the same best-fit values coming from the earlier \textbf{CMB+Quasar} analysis for generating the mock data. The input values of other 21~cm related free parameters (which did not appear in the \textbf{CMB+Quasar} analysis) are taken as $f^{\rm II}_{Xh}=0.2$ and $f^{ \rm II}_{ \alpha}=1.0$, consistent with the values estimated from the low redshift observations \citep{2006PhR...433..181F}.

Once the hypothetical 21~cm signal is constructed, we run an MCMC analysis to obtain the posterior distribution of all the nine parameters used in this analysis.

The constraints on the cosmological parameters are shown in Figure \ref{fig:all_pop2}, which allows a straightforward comparison with the previous \textbf{CMB+Quasar} analysis. Though the constraints on $H_0 , \Omega_{b}h^2 , \Omega_{c}h^2, n_s$ for the \textbf{CMB+Quasar+21cm$\mathbf{_{Pop-II}}$} are somewhat same as that of the \textbf{CMB+Quasar} analysis, the limits on $\sigma_{8}, A_{s}, \tau$ are tighter. For example, we note that the constraint on  $\tau$ is now significantly tighter in the \textbf{CMB+Quasar+21cm$\mathbf{_{Pop-II}}$} analysis. The $95\%$ C.L for $\tau$ is $\approx 0.005$ when 21~cm signal is added while it was $\approx 0.01$ earlier. This is a direct effect of including the $21~cm$ observations. The reason is that in the later redshifts (i.e., $z \leq 9$), the 21~cm signal starts probing the reionization history. This effectively means that by including the 21~cm observations we are now constraining reionization history using three different observations - CMB, Quasar and 21~cm signal. Hence the constraint on $\tau$ would be more stringent once the 21~cm signal is included. And, as it is well known that $\tau$ and $A_{s}$ (or $\sigma_8$) are correlated, the change in $\tau$ will also induce the change in the constraint of $\sigma_8$.

Interestingly, the fact that constraining $\tau$ from 21~cm experiments would lead to better limits on the cosmological parameters was already noted by  \cite{2016MNRAS.457.1864L} and \cite{2016PhRvD..93d3013L}. For example, in \cite{2016MNRAS.457.1864L}, the authors have used  combination of Planck observations, mock 21~cm power spectrum, and a mock global 21~cm signal to constrain both the astrophysical and cosmological parameters and found a tighter constraints on them. However, they consider only a small portion of the 21~cm signal (in the redshift range when the spin temperature is much larger compared to the background radiation temperature $T_{\gamma}$, i.e., $T_{s} \gg T_{\gamma}$). For this simplification, they did not have to introduce any extra parameters related to 21~cm signal. Also, unlike us, they did not consider any reionization related observations in their analysis.

The posterior distribution and the constraints on different parameters for the  \textbf{CMB+Quasar+21cm$\mathbf{_{Pop-II}}$} analysis are shown in Figure \ref{fig:triangle_21_pop2}. The correlation between the cosmological parameters are similar to what is expected from any standard analysis of the CMB data. The other interesting points of our analysis are as follows

\begin{enumerate}
    \item There is a strong anti correlation between $\sigma_8-f^{\mathrm{II}}_{Xh}$ and $\sigma_8-f^{\mathrm{II}}_{ \alpha}$. This is expected as we increase the value of $\sigma_8$, the structure formation will start earlier and the only way to keep the 21~cm signal unchanged is to decrease the efficiency of  Ly$-\alpha$ photon production (controlled by $f^{\mathrm{II}}_{ \alpha}$) and X-ray photon production (controlled by $f^{\mathrm{II}}_{Xh}$).
    
    \item We also note a strong positive correlation between $f^{\mathrm{II}}_{Xh}$ and 
    $f^{\mathrm{II}}_{ \alpha}$. This is also expected because increasing the Ly$-\alpha$ photon production efficiency will lead to a deeper absorption trough in the 21~cm signal and the only way to keep the minima of the signal unaltered is to increase the X-ray heating efficiency (which is determined by $\rm f^{II}_{Xh}$).
    
\end{enumerate}

We also show a comparison between the mock data (black line with error bars) and the 21~cm signal recovered from the MCMC analysis for this case in Figure \ref{fig:21cm_pop2}. The cyan shaded region shows the 1000 random samples from the MCMC chains and the purple curve corresponds to the best fit-values of the free parameters. It is evident from the figure that the recovered model matches quite well with the mock observation.

\subsection{CMB+Quasar+EDGES signal}

\begin{figure*}
    \centering
    \includegraphics[width=\textwidth]{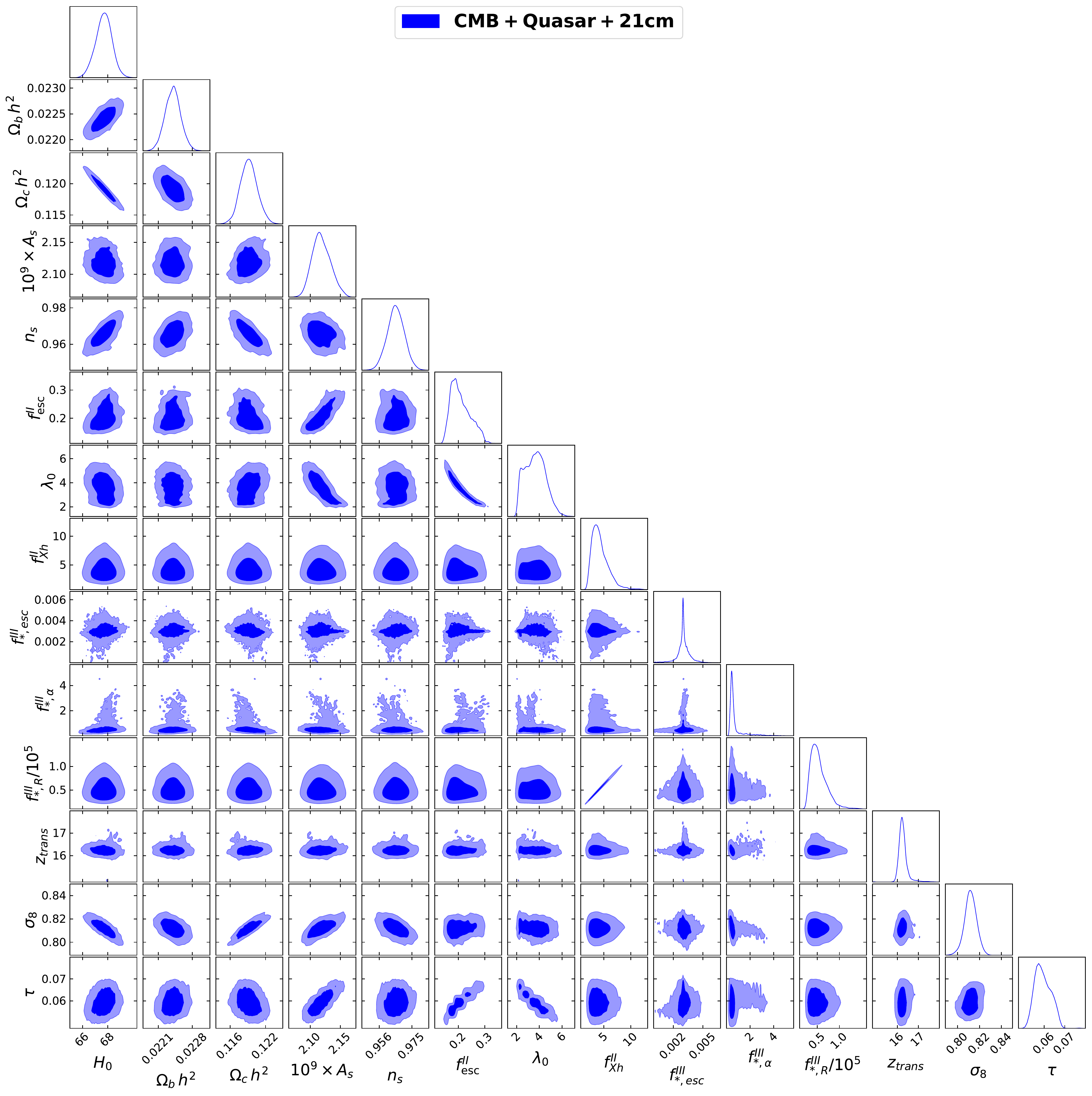}
    \caption{The marginalized posterior distribution of twelve free parameters and two derived parameters ($\sigma_8$ and $\tau$) obtained for the \textbf{CMB+Quasar+21cm} case. We show the $68\%$ and $95\%$ confidence contours in the two-dimensional plots. Note that the posteriors for the cosmological parameters are identical to those for the \textbf{CMB+Quasar+21cm} case shown in \fig{fig:all}. }
    \label{fig:triangle_21}
\end{figure*}

\begin{figure}
    \centering
    \includegraphics[width=0.5\textwidth]{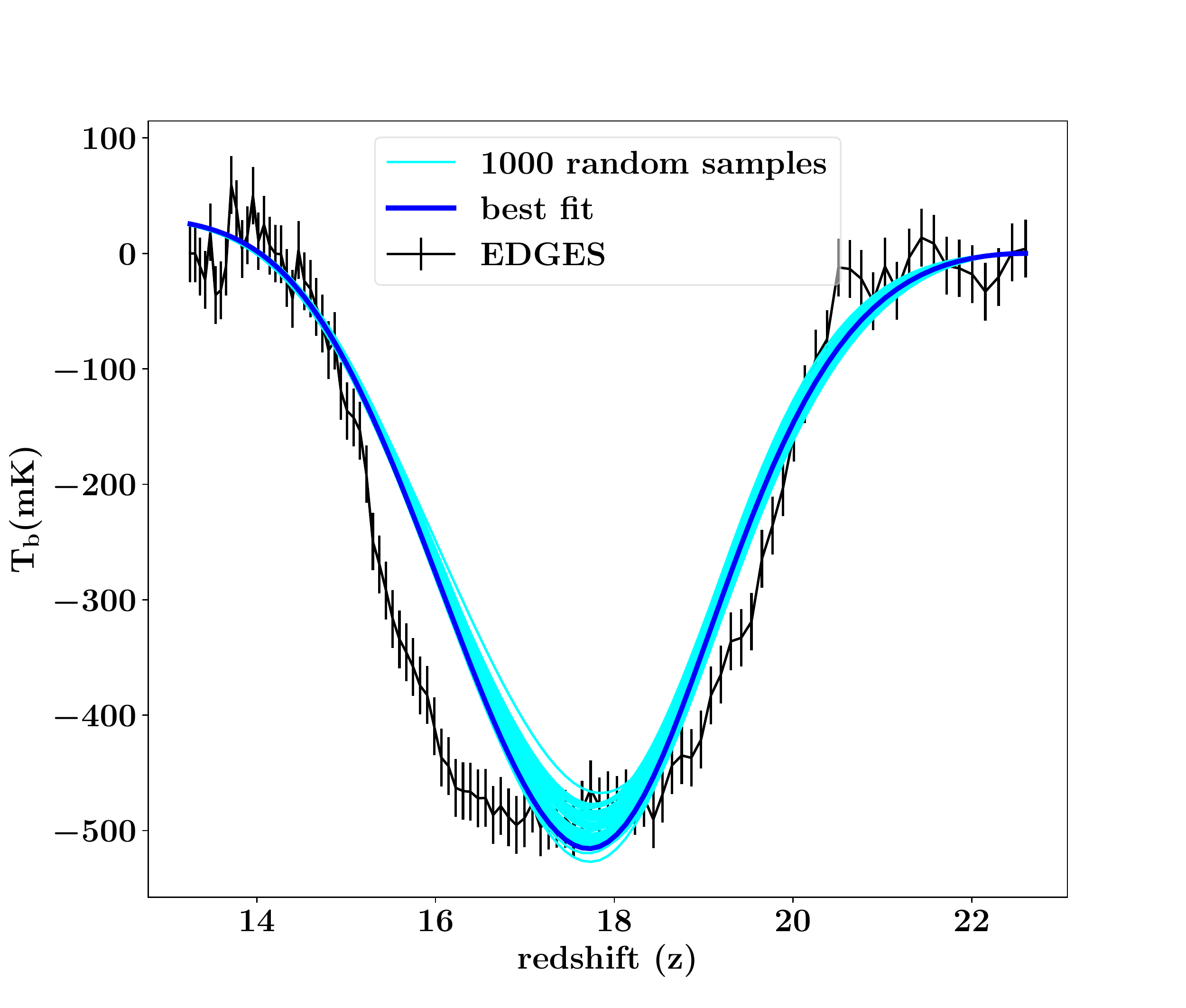}
    \caption{Comparison between EDGES observations \citep{2018Natur.555...67B} and the 21~cm signal recovered from our MCMC analysis in the \textbf{CMB+Quasar+21cm} case. The blue curve shows the best-fit model while as the cyan curves show 1000 random samples from the MCMC chains. The black line with the error bars represents the EDGES data points. Our model can match the position and depth of the EDGES absorption trough, however, matching the exact shape would require more complex evolution of the PopIII star formation than what is assumed in this paper.}
    \label{fig:21cm}
\end{figure}

Next, we describe the scenario where the EDGES 21~cm observation is included in our analysis. The basic analysis procedure for this scenario (\textbf{CMB+Quasar+21cm} hereafter) is very similar to that of \textbf{CMB+Quasar+21cm$_{\rm Pop-II}$}. The main differences are that we include the contribution of PopIII stars in \textbf{CMB+Quasar+21cm} analysis and an excess radio contribution has been considered here. As we will see, they are crucial in explaining the 21~cm data from EDGES. 

For the PopIII stars, the first point to note is that it is not straightforward to fix the value of star-forming efficiency $f_*^{\mathrm{III}}$ (recall that for PopII stars, we fixed it using the observations of the UV luminosity function at $z \sim 6 - 10$). However, this is not a serious obstacle as $f_*^{\mathrm{III}}$ appears as multiplicative factors with other free parameters and hence can be absorbed within them appropriately. Hence the free parameters for PopIII are
\begin{itemize}
    \item $f_{*, \mathrm{esc}}^{\mathrm{III}} \equiv f_*^{\mathrm{III}}~f_{\mathrm{esc}}^{\mathrm{IIII}}$: the escape of photons into the IGM.
    \item $f_{*, \alpha}^{\mathrm{III}} \equiv f_*^{\mathrm{III}}~f_{\alpha}^{\mathrm{III}}$: the Ly$\alpha$ flux.
    \item $f_{*, Xh}^{\mathrm{III}} \equiv f_*^{\mathrm{III}}~f_{Xh}^{\mathrm{III}}$: the heating efficiency of X-rays.
    \item $f_{*, R}^{\mathrm{II}} \equiv f_*^{\mathrm{III}}~f_R^{\mathrm{III}}$: the radio-SFR relation.
    \item $z_{\mathrm{trans}}$: the transition redshift for PopIII $\longrightarrow$ PopII stars.
\end{itemize}

Hence a joint analysis of all the data sets requires varying fifteen parameters (seven parameters of the earlier analyses and eight new parameters) simultaneously. This is an enormously challenging task which requires enormous computing time. We can simplify the analysis using some insights gained in our earlier work \citep{2020MNRAS.496.1445C}. It has been shown that it is sufficient to vary only the five parameters: $f_{Xh}^{\mathrm{II}}, f_{*, \mathrm{esc}}^{\mathrm{III}}, f_{*, \alpha}^{\mathrm{III}},  f^{\mathrm{III}}_{ *, R}, z_{\mathrm{trans}}$ in order to explain the EDGES signal. For the other four parameters related to the 21~cm signal, we assume the following: we fix $f^{\mathrm{II}}_{\alpha} = f^{\mathrm{II}}_R = 1$ which is equivalent to assuming that the Ly$\alpha$ and radio properties of the PopII stars remain similar to what is observed at lower redshifts. In addition, \citet{2020MNRAS.496.1445C} showed that the X-ray heating contribution of the PopIII stars is negligible compared to the PopII, so we take $f^{\mathrm{III}}_{Xh} = 0$. In \citet{2020MNRAS.496.1445C}, we found that the $z \leq 16$ part of the 21~cm signal is dominated by the X-ray heating where as the $z \geq 16$ is dominated by excess radio background and Lyman-$\alpha$ background. \citet{2020MNRAS.496.1445C} also suggest that for the  $z \leq 16$ part of the 21~cm signal, contribution comes mostly from PopII stars and not from PopIII stars. Hence, even if we consider X-ray contribution from PopIII, the match with the data would remain the same but at an expense of adding one more free parameter in the MCMC analysis.

So, the twelve free parameters for the joint analysis including the 21~cm signal are
\begin{equation*}
    \Theta =\{ H_{0}, \Omega_{b}h^{2}, \Omega_ch^{2}, A_{s}, n_s, f^{\mathrm{II}}_{\mathrm{esc}}, \lambda_{0}, f^{\mathrm{II}}_{Xh}, f^{\mathrm{III}}_{*, \mathrm{esc}},  f^{\mathrm{III}}_{*, \alpha}, f^{\mathrm{III}}_{*,R}, z_{\mathrm{trans}} \}
\end{equation*}

In the presence of the EDGES observations, the log-likelihood becomes
\begin{equation}
     \mathcal{L}=\mathcal{L}_{\mathrm{Pl}} + \mathcal{L}_{\mathrm{Re}} + \mathcal{L}_{21},
\end{equation}
where $\mathcal{L}_{21}$ is given by
\begin{equation}
    \mathcal{L}_{21}= \sum_i \left[\frac{\delta T_b^{\mathrm{obs}}(\nu_i) - \delta T_b^{\mathrm{th}}(\nu_i)}{\sigma_i}\right]^2,
\end{equation}
where $\delta T_b^{\mathrm{obs}}(\nu_i)$ and $\delta T_b^{\mathrm{th}}(\nu_i)$ are the 21~cm differential brightness temperatures from the observational data and theoretical model respectively and $\sigma_i$ are the observational errors. The sum is over all frequency channels $\nu_i$ of the EDGES data. We take $\sigma_{i} = 25$mK, consistent with \cite{2018Natur.564E..32H}. 

The posterior distributions of the different parameters are shown in \fig{fig:triangle_21}. The corresponding values can be found in Table~\ref{tab:table_1}. The first point to note is that the efficiency parameters related to the PopIII stars are required to be non-zero to match the data. We also find that the transition redshift $z_{\mathrm{trans}}$ is well constrained between $15.89$ and $16.68$, thus clearly indicating that the PopIII stars must cease to exist around $z \sim 16$. This is consistent with the findings of \citet{2020MNRAS.496.1445C}. Also, since the PopIII contribution terminates at such a high redshift, our model will automatically be less than the excess radio background allowed by the ARCADE-2 observations \citep{2011ApJ...734....5F}. Previously, other studies \cite{2020arXiv200804315R,2019MNRAS.487.3560C} too have found that the radio background must be turned off at $z \sim 16$. We require a large value of $f^{\mathrm{III}}_{*,R}$ ($\approx 10^4$) suggesting a highly efficient radio emission from the PopIII stars. Interestingly, \cite{2020arXiv200804315R} have also found a similar value for this quantity while studying the effect of inhomogeneous radio background (coming from a population of high-redshift galaxies) in both 21~cm power spectrum and the global signal. 

Another interesting feature to note from this figure is a very strong correlation between $f^{\mathrm{III}}_{*,R}$ and $f^{\mathrm{II}}_{Xh}$. This is expected because higher the radio background, deeper will be the absorption trough of the 21~cm signal around $z_{\mathrm{trans}}$. However, the observed signal approaches zero at redshift $z \sim 14$ and the only way to match with this feature is to add more X-ray heating.

As far as the reionization history is concerned, we find that $\tau$ is correlated with $f^{\mathrm{II}}_{\mathrm{esc}}$ and is anti-correlated with $\lambda_0$, similar to the \textbf{CMB+Quasar} analysis. However we see that there is no correlation between $\tau$ and $f^{\mathrm{III}}_{*, \mathrm{esc}}$, thus implying that the reionization history is not sensitive to the details of the PopIII stars. There are some differences in the posterior distributions of $H_0$, $\Omega_b h^2$ and $\Omega_c h^2$, however, these are not statistically significant at this stage.

In \fig{fig:all}, we compare the constraints on the cosmological parameters for the \textbf{CMB+Quasar+21cm} with the earlier analyses \textbf{only-CMB} and \textbf{CMB+Quasar}. As can be seen, the constraints for the \textbf{CMB+Quasar+21cm} are almost identical to the \textbf{CMB+Quasar}, thus implying that the EDGES data do not have any significant effect on the cosmological parameters of the $\Lambda$CDM model. The \textbf{CMB+Quasar+21cm} prefers slightly higher values of $\tau$ compared to \textbf{only-CMB} analysis, driven by the presence of PopIII stars.

The redshift evolution of different reionization-related quantities has been compared with the \textbf{CMB+Quasar} in \fig{fig:redshift_reionization}. The dashed blue lines show the best-fit results of \textbf{CMB+Quasar+21cm} analysis whereas the cyan regions correspond to 1000 random samples taken from the MCMC chains. We find that the \textbf{CMB+Quasar+21cm} analysis prefers somewhat earlier reionization which is the result of including the PopIII stars in the model. We also find that the \textbf{CMB+Quasar+21cm} allows a wider range of reionization histories compared to \textbf{CMB+Quasar} which too is a consequence of including additional free parameters. The best-fit models for the two cases, however, are almost identical which can also be seen from the parameter values in Table~\ref{tab:table_1} along with the corresponding log-likelihood values.

Finally, we show the comparison between the EDGES data and the 21~cm signal obtained from the MCMC analysis in \fig{fig:21cm}. The cyan shaded region shows 1000 random samples from the MCMC chains and the blue curve corresponds to the best-fit values of the free parameters. Though the amplitude of the absorption matches with the observations, our theoretical model cannot produce the ``flat-bottom'' shape as seen in the observed signal. \cite{2020MNRAS.496.1445C} has shown that to reproduce this shape, one needs to employ a much more complicated shape for the PopIII star formation, e.g., a fourth-order log-polynomial. This would naturally lead to a significantly more model parameters which we postpone to a later work.

\section{Conclusion and Discussion}
\label{section:conclusion}
In this work, we introduce \texttt{CosmoReionMC}, an advanced Markov Chain Monte Carlo (MCMC) based parameter estimation package using a variety of cosmological and reionization-related observations. In particular, we include the 21~cm global signal from cosmic dawn in the analysis. To our knowledge, this is the first time the 21~cm data has been combined with the cosmological observations.

Our formalism is based on (i) a reionization model developed by \citet{2005MNRAS.361..577C,2006MNRAS.371L..55C} with improvements by \citet{2011MNRAS.413.1569M}, (ii) a model for 21~cm signal based on \citet{2006MNRAS.371..867F,2019MNRAS.482.1456C} and (iii) a publicly available package \texttt{CAMB} \citep{Lewis:2013hha} for computing the CMB anisotropies, modified to account for arbitrary reionization histories. The code is then coupled to a MCMC code \texttt{emcee} \citep{2013PASP..125..306F} to compute the posteriors of the free parameters.

Depending on the data sets used, we perform three different analyses with our package:

\begin{enumerate}
    \item \textbf{only-CMB:} We use only Planck2018 CMB anisotropy data in this analysis. Compared to the conventional analyses of the CMB data, this is equivalent to replacing the paramterized reionization history (either tanh or more advanced parametrization) with a physically motivated reionization model. We find that the bounds on the cosmological parameters are similar to what is obtained in the conventional tanh parameterization of reionization history.
    
    \item \textbf{CMB+Quasar:} In this analysis we include the reionization-related observations based on quasar absorption spectra, namely, the constraints on the hydrogen photoionization rate \citep{2011MNRAS.412.2543C,2013MNRAS.436.1023B,2018MNRAS.473..560D}, the redshift distribution of the Lyman-limit systems \citep{2010ApJ...721.1448S,2019MNRAS.482.1456C}, the model-independent constraints using the dark pixels \citep{2015MNRAS.447..499M} and prior on the end of reionization based on the Ly$\alpha$ opacity fluctuations \citep{2019MNRAS.485L..24K,2020arXiv200308958C}. The inclusion of these observations lead to much tighter constraints on $\tau$ compared to \textbf{only-CMB} analysis which, in turn, lead to tighter constraints on the amplitude of the primordial power spectrum $A_s$ and $\sigma_8$.
    
    \item \textbf{CMB+Quasar+21cm:} In this case we add the EDGES observations of 21~cm global signal \citep{2018Natur.555...67B} along with the CMB and quasar absorption data. The main difference of this analysis with the earlier ones is that the EDGES data requires a radio-efficient short-lived population of metal-free (PopIII) stars which give rise to the absorption trough. The inclusion of the PopIII stars allow reionization histories that start earlier than what is found in the \textbf{CMB+Quasar} analysis. With the present EDGES data, the constraints on the cosmological parameters are very similar in the two cases.
\end{enumerate}

There are a couple of caveats worth highlighting here. Our analysis assumes all the efficiency parameters related to the PopII and PopIII stars to be redshift-independent, which is clearly too simplistic. In fact, several studies suggest that they are possibly function of halo masses and redshifts. For the reionization-related data, the only parameter which is of relevance is $N_{\mathrm{ion}}$, see \eqn{eq:Nion}, which is a multiplication of several physical parameters including $f_*$ and $f_{\mathrm{esc}}$. In such cases, it is possible to treat $N_{\mathrm{ion}}$ as an arbitrary function of $z$ and constrain its evolution using Principal Component Analysis \citep{2011MNRAS.413.1569M}. Similarly, one can also include halo mass-dependence using some simple power-law parameterization \citep[see e.g.,][]{2019MNRAS.490.2855Y, 2019MNRAS.486.2215K, 2015MNRAS.451.2544P, 2013MNRAS.431.2826F}. Inclusion of the 21~cm data, however, introduces additional efficiency parameters related to the Ly$\alpha$ radiation, X-ray heating and the radio background. In such a scenario, implementing $z$ and $M$-dependent efficiency parameters would lead to a significantly larger number of free parameters thus making the parameter estimation exceedingly challenging. Such issues will be explored elsewhere.

The second point to keep in mind is that the origin and amplitude of the EDGES 21~cm signal is still highly debated \citep[see, e.g.,][]{2018Natur.564E..32H,2020MNRAS.492...22S}. In case the amplitude of the signal is found to be less than what is currently claimed, the need for an excess radio background and hence the need for introducing PopIII stars in the model will go away. Whether this will have any effect on the cosmological parameters needs to be investigated.

The \texttt{CosmoReionMC} package can be extended to study different kinds of problems. For example, it is straightforward to include other cosmological data sets (say, BAO) in the analysis. On the astrophysical side too, we can include other constraints on the reionization history. The package can also be used to probing non-standard extensions to the standard $\Lambda$CDM model, e.g., evolving dark energy and light dark matter particles. Similarly, newer measurements of the 21~cm global signal are expected from several other experiments aiming to detect it, e.g., the Shaped Antenna measurement of the background RAdio Spectrum \citep[SARAS,][]{Patra_2013,Singh_2017}, the Large-Aperture Experiment to Detect the Dark Ages \citep[LEDA,][]{Greenhill_2012}, SCI-HI \citep{Voytek_2014}, the Broadband Instrument for Global Hydrogen Reionisation Signal \citep[BIGHORNS,][]{Sokolowski_2015}, and the Cosmic Twilight Polarimeter \citep[CTP,][]{Nahn_2018}. It would be interesting to check if these measurements have any impact on the cosmological and reionization parameters. Such possibilities will be explored in the future. In addition, we are also planning to make \texttt{CosmoReionMC} public in the near future.

To summarise, the \texttt{CosmoReionMC} is a unique parameter estimation package which can probe cosmological and astrophysical parameters related to reionization simultaneously using a wide variety of data sets. This should have many interesting applications in the future.

\section*{Acknowledgements}
We thank the anonymous referee for their helpful and constructive comments on the first version of the paper. AC and TRC acknowledge support of the Department of Atomic Energy, Government of India, under project no.~12-R\&D-TFR-5.02-0700. TRC is also supported by the Associateship Scheme of ICTP.

\section*{Data Availability}

The observational data used here are taken from literature and the code underlying this article will be shared on reasonable request to the corresponding author.


\bibliographystyle{mnras}
\bibliography{main_ver2_revised}

\begin{thebibliography}{}
\makeatletter
\relax
\def\mn@urlcharsother{\let\do\@makeother \do\$\do\&\do\#\do\^\do\_\do\%\do\~}
\def\mn@doi{\begingroup\mn@urlcharsother \@ifnextchar [ {\mn@doi@}
  {\mn@doi@[]}}
\def\mn@doi@[#1]#2{\def\@tempa{#1}\ifx\@tempa\@empty \href
  {http://dx.doi.org/#2} {doi:#2}\else \href {http://dx.doi.org/#2} {#1}\fi
  \endgroup}
\def\mn@eprint#1#2{\mn@eprint@#1:#2::\@nil}
\def\mn@eprint@arXiv#1{\href {http://arxiv.org/abs/#1} {{\tt arXiv:#1}}}
\def\mn@eprint@dblp#1{\href {http://dblp.uni-trier.de/rec/bibtex/#1.xml}
  {dblp:#1}}
\def\mn@eprint@#1:#2:#3:#4\@nil{\def\@tempa {#1}\def\@tempb {#2}\def\@tempc
  {#3}\ifx \@tempc \@empty \let \@tempc \@tempb \let \@tempb \@tempa \fi \ifx
  \@tempb \@empty \def\@tempb {arXiv}\fi \@ifundefined
  {mn@eprint@\@tempb}{\@tempb:\@tempc}{\expandafter \expandafter \csname
  mn@eprint@\@tempb\endcsname \expandafter{\@tempc}}}

\bibitem[\protect\citeauthoryear{{Akeret}, {Seehars}, {Amara}, {Refregier}  \&
  {Csillaghy}}{{Akeret} et~al.}{2013}]{2013A&C.....2...27A}
{Akeret} J.,  {Seehars} S.,  {Amara} A.,  {Refregier} A.,   {Csillaghy} A.,
  2013, \mn@doi [Astronomy and Computing] {10.1016/j.ascom.2013.06.003}, \href
  {https://ui.adsabs.harvard.edu/abs/2013A&C.....2...27A} {2, 27}

\bibitem[\protect\citeauthoryear{{Barkana} \& {Loeb}}{{Barkana} \&
  {Loeb}}{2001}]{2001PhR...349..125B}
{Barkana} R.,  {Loeb} A.,  2001, \mn@doi [\physrep]
  {10.1016/S0370-1573(01)00019-9}, \href
  {https://ui.adsabs.harvard.edu/abs/2001PhR...349..125B} {349, 125}

\bibitem[\protect\citeauthoryear{{Barkana}, {Outmezguine}, {Redigol}  \&
  {Volansky}}{{Barkana} et~al.}{2018}]{2018PhRvD..98j3005B}
{Barkana} R.,  {Outmezguine} N.~J.,  {Redigol} D.,   {Volansky} T.,  2018,
  \mn@doi [\prd] {10.1103/PhysRevD.98.103005}, \href
  {https://ui.adsabs.harvard.edu/abs/2018PhRvD..98j3005B} {98, 103005}

\bibitem[\protect\citeauthoryear{{Becker} \& {Bolton}}{{Becker} \&
  {Bolton}}{2013}]{2013MNRAS.436.1023B}
{Becker} G.~D.,  {Bolton} J.~S.,  2013, \mn@doi [\mnras]
  {10.1093/mnras/stt1610}, \href
  {https://ui.adsabs.harvard.edu/abs/2013MNRAS.436.1023B} {436, 1023}

\bibitem[\protect\citeauthoryear{{Becker}, {Bolton}, {Madau}, {Pettini},
  {Ryan-Weber}  \& {Venemans}}{{Becker} et~al.}{2015}]{2015MNRAS.447.3402B}
{Becker} G.~D.,  {Bolton} J.~S.,  {Madau} P.,  {Pettini} M.,  {Ryan-Weber}
  E.~V.,   {Venemans} B.~P.,  2015, \mn@doi [\mnras] {10.1093/mnras/stu2646},
  \href {https://ui.adsabs.harvard.edu/abs/2015MNRAS.447.3402B} {447, 3402}

\bibitem[\protect\citeauthoryear{{Bosman}, {Fan}, {Jiang}, {Reed}, {Matsuoka},
  {Becker}  \& {Haehnelt}}{{Bosman} et~al.}{2018}]{2018MNRAS.479.1055B}
{Bosman} S. E.~I.,  {Fan} X.,  {Jiang} L.,  {Reed} S.,  {Matsuoka} Y.,
  {Becker} G.,   {Haehnelt} M.,  2018, \mn@doi [\mnras]
  {10.1093/mnras/sty1344}, \href
  {https://ui.adsabs.harvard.edu/abs/2018MNRAS.479.1055B} {479, 1055}

\bibitem[\protect\citeauthoryear{{Bouwens}, {Illingworth}, {Blakeslee}  \&
  {Franx}}{{Bouwens} et~al.}{2006}]{2006ApJ...653...53B}
{Bouwens} R.~J.,  {Illingworth} G.~D.,  {Blakeslee} J.~P.,   {Franx} M.,  2006,
  \mn@doi [\apj] {10.1086/498733}, \href
  {https://ui.adsabs.harvard.edu/abs/2006ApJ...653...53B} {653, 53}

\bibitem[\protect\citeauthoryear{{Bouwens} et~al.,}{{Bouwens}
  et~al.}{2015}]{2015ApJ...803...34B}
{Bouwens} R.~J.,  et~al., 2015, \mn@doi [\apj] {10.1088/0004-637X/803/1/34},
  \href {https://ui.adsabs.harvard.edu/abs/2015ApJ...803...34B} {803, 34}

\bibitem[\protect\citeauthoryear{{Bowman}, {Rogers}, {Monsalve}, {Mozdzen}  \&
  {Mahesh}}{{Bowman} et~al.}{2018}]{2018Natur.555...67B}
{Bowman} J.~D.,  {Rogers} A. E.~E.,  {Monsalve} R.~A.,  {Mozdzen} T.~J.,
  {Mahesh} N.,  2018, \mn@doi [\nat] {10.1038/nature25792}, \href
  {https://ui.adsabs.harvard.edu/abs/2018Natur.555...67B} {555, 67}

\bibitem[\protect\citeauthoryear{{Bradley}, {Tauscher}, {Rapetti}  \&
  {Burns}}{{Bradley} et~al.}{2019}]{2019ApJ...874..153B}
{Bradley} R.~F.,  {Tauscher} K.,  {Rapetti} D.,   {Burns} J.~O.,  2019, \mn@doi
  [\apj] {10.3847/1538-4357/ab0d8b}, \href
  {https://ui.adsabs.harvard.edu/abs/2019ApJ...874..153B} {874, 153}

\bibitem[\protect\citeauthoryear{{Bromm} \& {Loeb}}{{Bromm} \&
  {Loeb}}{2003}]{2003Natur.425..812B}
{Bromm} V.,  {Loeb} A.,  2003, \mn@doi [\nat] {10.1038/nature02071}, \href
  {https://ui.adsabs.harvard.edu/abs/2003Natur.425..812B} {425, 812}

\bibitem[\protect\citeauthoryear{{Bruzual} \& {Charlot}}{{Bruzual} \&
  {Charlot}}{2003}]{2003MNRAS.344.1000B}
{Bruzual} G.,  {Charlot} S.,  2003, \mn@doi [\mnras]
  {10.1046/j.1365-8711.2003.06897.x}, \href
  {https://ui.adsabs.harvard.edu/abs/2003MNRAS.344.1000B} {344, 1000}

\bibitem[\protect\citeauthoryear{{Calverley}, {Becker}, {Haehnelt}  \&
  {Bolton}}{{Calverley} et~al.}{2011}]{2011MNRAS.412.2543C}
{Calverley} A.~P.,  {Becker} G.~D.,  {Haehnelt} M.~G.,   {Bolton} J.~S.,  2011,
  \mn@doi [\mnras] {10.1111/j.1365-2966.2010.18072.x}, \href
  {https://ui.adsabs.harvard.edu/abs/2011MNRAS.412.2543C} {412, 2543}

\bibitem[\protect\citeauthoryear{{Chatterjee}, {Dayal}, {Choudhury}  \&
  {Hutter}}{{Chatterjee} et~al.}{2019}]{2019MNRAS.487.3560C}
{Chatterjee} A.,  {Dayal} P.,  {Choudhury} T.~R.,   {Hutter} A.,  2019, \mn@doi
  [\mnras] {10.1093/mnras/stz1444}, \href
  {https://ui.adsabs.harvard.edu/abs/2019MNRAS.487.3560C} {487, 3560}

\bibitem[\protect\citeauthoryear{{Chatterjee}, {Dayal}, {Choudhury}  \&
  {Schneider}}{{Chatterjee} et~al.}{2020}]{2020MNRAS.496.1445C}
{Chatterjee} A.,  {Dayal} P.,  {Choudhury} T.~R.,   {Schneider} R.,  2020,
  \mn@doi [\mnras] {10.1093/mnras/staa1609}, \href
  {https://ui.adsabs.harvard.edu/abs/2020MNRAS.496.1445C} {496, 1445}

\bibitem[\protect\citeauthoryear{{Chiu} \& {Ostriker}}{{Chiu} \&
  {Ostriker}}{2000}]{2000ApJ...534..507C}
{Chiu} W.~A.,  {Ostriker} J.~P.,  2000, \mn@doi [\apj] {10.1086/308780}, \href
  {https://ui.adsabs.harvard.edu/abs/2000ApJ...534..507C} {534, 507}

\bibitem[\protect\citeauthoryear{{Choudhury}}{{Choudhury}}{2009}]{2009CSci...97..841C}
{Choudhury} T.~R.,  2009, Current Science, \href
  {https://ui.adsabs.harvard.edu/abs/2009CSci...97..841C} {97, 841}

\bibitem[\protect\citeauthoryear{{Choudhury} \& {Ferrara}}{{Choudhury} \&
  {Ferrara}}{2005}]{2005MNRAS.361..577C}
{Choudhury} T.~R.,  {Ferrara} A.,  2005, \mn@doi [\mnras]
  {10.1111/j.1365-2966.2005.09196.x}, \href
  {https://ui.adsabs.harvard.edu/abs/2005MNRAS.361..577C} {361, 577}

\bibitem[\protect\citeauthoryear{{Choudhury} \& {Ferrara}}{{Choudhury} \&
  {Ferrara}}{2006}]{2006MNRAS.371L..55C}
{Choudhury} T.~R.,  {Ferrara} A.,  2006, \mn@doi [\mnras]
  {10.1111/j.1745-3933.2006.00207.x}, \href
  {https://ui.adsabs.harvard.edu/abs/2006MNRAS.371L..55C} {371, L55}

\bibitem[\protect\citeauthoryear{{Choudhury}, {Paranjape}  \&
  {Bosman}}{{Choudhury} et~al.}{2020}]{2020arXiv200308958C}
{Choudhury} T.~R.,  {Paranjape} A.,   {Bosman} S. E.~I.,  2020, arXiv e-prints,
  \href {https://ui.adsabs.harvard.edu/abs/2020arXiv200308958C} {p.
  arXiv:2003.08958}

\bibitem[\protect\citeauthoryear{{Crighton}, {Prochaska}, {Murphy}, {O'Meara},
  {Worseck}  \& {Smith}}{{Crighton} et~al.}{2019}]{2019MNRAS.482.1456C}
{Crighton} N. H.~M.,  {Prochaska} J.~X.,  {Murphy} M.~T.,  {O'Meara} J.~M.,
  {Worseck} G.,   {Smith} B.~D.,  2019, \mn@doi [\mnras]
  {10.1093/mnras/sty2762}, \href
  {https://ui.adsabs.harvard.edu/abs/2019MNRAS.482.1456C} {482, 1456}

\bibitem[\protect\citeauthoryear{{D'Aloisio}, {McQuinn}, {Davies}  \&
  {Furlanetto}}{{D'Aloisio} et~al.}{2018}]{2018MNRAS.473..560D}
{D'Aloisio} A.,  {McQuinn} M.,  {Davies} F.~B.,   {Furlanetto} S.~R.,  2018,
  \mn@doi [\mnras] {10.1093/mnras/stx2341}, \href
  {https://ui.adsabs.harvard.edu/abs/2018MNRAS.473..560D} {473, 560}

\bibitem[\protect\citeauthoryear{{Davies}, {Mutch}, {Qin}, {Mesinger}, {Poole}
  \& {Wyithe}}{{Davies} et~al.}{2019}]{2019MNRAS.489..977D}
{Davies} J.~E.,  {Mutch} S.~J.,  {Qin} Y.,  {Mesinger} A.,  {Poole} G.~B.,
  {Wyithe} J. S.~B.,  2019, \mn@doi [\mnras] {10.1093/mnras/stz2241}, \href
  {https://ui.adsabs.harvard.edu/abs/2019MNRAS.489..977D} {489, 977}

\bibitem[\protect\citeauthoryear{{Dayal} \& {Ferrara}}{{Dayal} \&
  {Ferrara}}{2018}]{2018PhR...780....1D}
{Dayal} P.,  {Ferrara} A.,  2018, \mn@doi [\physrep]
  {10.1016/j.physrep.2018.10.002}, \href
  {https://ui.adsabs.harvard.edu/abs/2018PhR...780....1D} {780, 1}

\bibitem[\protect\citeauthoryear{{Douspis}, {Aghanim}, {Ili{\'c}}  \&
  {Langer}}{{Douspis} et~al.}{2015}]{2015A&A...580L...4D}
{Douspis} M.,  {Aghanim} N.,  {Ili{\'c}} S.,   {Langer} M.,  2015, \mn@doi
  [\aap] {10.1051/0004-6361/201526543}, \href
  {https://ui.adsabs.harvard.edu/abs/2015A&A...580L...4D} {580, L4}

\bibitem[\protect\citeauthoryear{{Eilers}, {Davies}, {Hennawi}, {Prochaska},
  {Luki{\'c}}  \& {Mazzucchelli}}{{Eilers} et~al.}{2017}]{2017ApJ...840...24E}
{Eilers} A.-C.,  {Davies} F.~B.,  {Hennawi} J.~F.,  {Prochaska} J.~X.,
  {Luki{\'c}} Z.,   {Mazzucchelli} C.,  2017, \mn@doi [\apj]
  {10.3847/1538-4357/aa6c60}, \href
  {https://ui.adsabs.harvard.edu/abs/2017ApJ...840...24E} {840, 24}

\bibitem[\protect\citeauthoryear{{Eilers}, {Davies}  \& {Hennawi}}{{Eilers}
  et~al.}{2018}]{2018ApJ...864...53E}
{Eilers} A.-C.,  {Davies} F.~B.,   {Hennawi} J.~F.,  2018, \mn@doi [\apj]
  {10.3847/1538-4357/aad4fd}, \href
  {https://ui.adsabs.harvard.edu/abs/2018ApJ...864...53E} {864, 53}

\bibitem[\protect\citeauthoryear{{Ewall-Wice}, {Chang}, {Lazio}, {Dor{\'e}},
  {Seiffert}  \& {Monsalve}}{{Ewall-Wice} et~al.}{2018}]{2018ApJ...868...63E}
{Ewall-Wice} A.,  {Chang} T.~C.,  {Lazio} J.,  {Dor{\'e}} O.,  {Seiffert} M.,
  {Monsalve} R.~A.,  2018, \mn@doi [\apj] {10.3847/1538-4357/aae51d}, \href
  {https://ui.adsabs.harvard.edu/abs/2018ApJ...868...63E} {868, 63}

\bibitem[\protect\citeauthoryear{{Ewall-Wice}, {Chang}  \&
  {Lazio}}{{Ewall-Wice} et~al.}{2020}]{2020MNRAS.492.6086E}
{Ewall-Wice} A.,  {Chang} T.-C.,   {Lazio} T. J.~W.,  2020, \mn@doi [\mnras]
  {10.1093/mnras/stz3501}, \href
  {https://ui.adsabs.harvard.edu/abs/2020MNRAS.492.6086E} {492, 6086}

\bibitem[\protect\citeauthoryear{{Feng} \& {Holder}}{{Feng} \&
  {Holder}}{2018}]{2018ApJ...858L..17F}
{Feng} C.,  {Holder} G.,  2018, \mn@doi [\apjl] {10.3847/2041-8213/aac0fe},
  \href {https://ui.adsabs.harvard.edu/abs/2018ApJ...858L..17F} {858, L17}

\bibitem[\protect\citeauthoryear{{Ferrara} \& {Loeb}}{{Ferrara} \&
  {Loeb}}{2013}]{2013MNRAS.431.2826F}
{Ferrara} A.,  {Loeb} A.,  2013, \mn@doi [\mnras] {10.1093/mnras/stt381}, \href
  {https://ui.adsabs.harvard.edu/abs/2013MNRAS.431.2826F} {431, 2826}

\bibitem[\protect\citeauthoryear{{Fialkov} \& {Barkana}}{{Fialkov} \&
  {Barkana}}{2019}]{2019MNRAS.486.1763F}
{Fialkov} A.,  {Barkana} R.,  2019, \mn@doi [\mnras] {10.1093/mnras/stz873},
  \href {https://ui.adsabs.harvard.edu/abs/2019MNRAS.486.1763F} {486, 1763}

\bibitem[\protect\citeauthoryear{{Fialkov}, {Barkana}, {Visbal},
  {Tseliakhovich}  \& {Hirata}}{{Fialkov} et~al.}{2013}]{2013MNRAS.432.2909F}
{Fialkov} A.,  {Barkana} R.,  {Visbal} E.,  {Tseliakhovich} D.,   {Hirata}
  C.~M.,  2013, \mn@doi [\mnras] {10.1093/mnras/stt650}, \href
  {https://ui.adsabs.harvard.edu/abs/2013MNRAS.432.2909F} {432, 2909}

\bibitem[\protect\citeauthoryear{{Field}}{{Field}}{1958}]{1958PIRE...46..240F}
{Field} G.~B.,  1958, \mn@doi [Proceedings of the IRE]
  {10.1109/JRPROC.1958.286741}, \href
  {https://ui.adsabs.harvard.edu/abs/1958PIRE...46..240F} {46, 240}

\bibitem[\protect\citeauthoryear{{Fixsen} et~al.,}{{Fixsen}
  et~al.}{2011}]{2011ApJ...734....5F}
{Fixsen} D.~J.,  et~al., 2011, \mn@doi [\apj] {10.1088/0004-637X/734/1/5},
  \href {https://ui.adsabs.harvard.edu/abs/2011ApJ...734....5F} {734, 5}

\bibitem[\protect\citeauthoryear{{Foreman-Mackey}, {Hogg}, {Lang}  \&
  {Goodman}}{{Foreman-Mackey} et~al.}{2013}]{2013PASP..125..306F}
{Foreman-Mackey} D.,  {Hogg} D.~W.,  {Lang} D.,   {Goodman} J.,  2013, \mn@doi
  [\pasp] {10.1086/670067}, \href
  {https://ui.adsabs.harvard.edu/abs/2013PASP..125..306F} {125, 306}

\bibitem[\protect\citeauthoryear{{Fraser} et~al.,}{{Fraser}
  et~al.}{2018}]{2018PhLB..785..159F}
{Fraser} S.,  et~al., 2018, \mn@doi [Physics Letters B]
  {10.1016/j.physletb.2018.08.035}, \href
  {https://ui.adsabs.harvard.edu/abs/2018PhLB..785..159F} {785, 159}

\bibitem[\protect\citeauthoryear{{Fumagalli}, {O'Meara}, {Prochaska}  \&
  {Worseck}}{{Fumagalli} et~al.}{2013}]{2013ApJ...775...78F}
{Fumagalli} M.,  {O'Meara} J.~M.,  {Prochaska} J.~X.,   {Worseck} G.,  2013,
  \mn@doi [\apj] {10.1088/0004-637X/775/1/78}, \href
  {https://ui.adsabs.harvard.edu/abs/2013ApJ...775...78F} {775, 78}

\bibitem[\protect\citeauthoryear{{Furlanetto}}{{Furlanetto}}{2006}]{2006MNRAS.371..867F}
{Furlanetto} S.~R.,  2006, \mn@doi [\mnras] {10.1111/j.1365-2966.2006.10725.x},
  \href {https://ui.adsabs.harvard.edu/abs/2006MNRAS.371..867F} {371, 867}

\bibitem[\protect\citeauthoryear{{Furlanetto}, {Oh}  \& {Briggs}}{{Furlanetto}
  et~al.}{2006}]{2006PhR...433..181F}
{Furlanetto} S.~R.,  {Oh} S.~P.,   {Briggs} F.~H.,  2006, \mn@doi [\physrep]
  {10.1016/j.physrep.2006.08.002}, \href
  {https://ui.adsabs.harvard.edu/abs/2006PhR...433..181F} {433, 181}

\bibitem[\protect\citeauthoryear{{Ghara} \& {Mellema}}{{Ghara} \&
  {Mellema}}{2020}]{2020MNRAS.492..634G}
{Ghara} R.,  {Mellema} G.,  2020, \mn@doi [\mnras] {10.1093/mnras/stz3513},
  \href {https://ui.adsabs.harvard.edu/abs/2020MNRAS.492..634G} {492, 634}

\bibitem[\protect\citeauthoryear{{Ghara}, {Giri}, {Ciardi}, {Mellema}  \&
  {Zaroubi}}{{Ghara} et~al.}{2021}]{2021MNRAS.503.4551G}
{Ghara} R.,  {Giri} S.~K.,  {Ciardi} B.,  {Mellema} G.,   {Zaroubi} S.,  2021,
  \mn@doi [\mnras] {10.1093/mnras/stab776}, \href
  {https://ui.adsabs.harvard.edu/abs/2021MNRAS.503.4551G} {503, 4551}

\bibitem[\protect\citeauthoryear{{Gillet}, {Mesinger}, {Greig}, {Liu}  \&
  {Ucci}}{{Gillet} et~al.}{2019}]{2019MNRAS.484..282G}
{Gillet} N.,  {Mesinger} A.,  {Greig} B.,  {Liu} A.,   {Ucci} G.,  2019,
  \mn@doi [\mnras] {10.1093/mnras/stz010}, \href
  {https://ui.adsabs.harvard.edu/abs/2019MNRAS.484..282G} {484, 282}

\bibitem[\protect\citeauthoryear{{Greenhill} \& {Bernardi}}{{Greenhill} \&
  {Bernardi}}{2012}]{Greenhill_2012}
{Greenhill} L.~J.,  {Bernardi} G.,  2012, arXiv e-prints, \href
  {https://ui.adsabs.harvard.edu/abs/2012arXiv1201.1700G} {p. arXiv:1201.1700}

\bibitem[\protect\citeauthoryear{{Greig} \& {Mesinger}}{{Greig} \&
  {Mesinger}}{2015}]{2015MNRAS.449.4246G}
{Greig} B.,  {Mesinger} A.,  2015, \mn@doi [\mnras] {10.1093/mnras/stv571},
  \href {https://ui.adsabs.harvard.edu/abs/2015MNRAS.449.4246G} {449, 4246}

\bibitem[\protect\citeauthoryear{{Greig} \& {Mesinger}}{{Greig} \&
  {Mesinger}}{2017a}]{2017MNRAS.465.4838G}
{Greig} B.,  {Mesinger} A.,  2017a, \mn@doi [\mnras] {10.1093/mnras/stw3026},
  \href {https://ui.adsabs.harvard.edu/abs/2017MNRAS.465.4838G} {465, 4838}

\bibitem[\protect\citeauthoryear{{Greig} \& {Mesinger}}{{Greig} \&
  {Mesinger}}{2017b}]{2017MNRAS.472.2651G}
{Greig} B.,  {Mesinger} A.,  2017b, \mn@doi [\mnras] {10.1093/mnras/stx2118},
  \href {https://ui.adsabs.harvard.edu/abs/2017MNRAS.472.2651G} {472, 2651}

\bibitem[\protect\citeauthoryear{{Greig} \& {Mesinger}}{{Greig} \&
  {Mesinger}}{2018}]{2018MNRAS.477.3217G}
{Greig} B.,  {Mesinger} A.,  2018, \mn@doi [\mnras] {10.1093/mnras/sty796},
  \href {https://ui.adsabs.harvard.edu/abs/2018MNRAS.477.3217G} {477, 3217}

\bibitem[\protect\citeauthoryear{{Greig}, {Mesinger}  \& {Koopmans}}{{Greig}
  et~al.}{2020}]{2020MNRAS.491.1398G}
{Greig} B.,  {Mesinger} A.,   {Koopmans} L. V.~E.,  2020, \mn@doi [\mnras]
  {10.1093/mnras/stz3138}, \href
  {https://ui.adsabs.harvard.edu/abs/2020MNRAS.491.1398G} {491, 1398}

\bibitem[\protect\citeauthoryear{{Haiman}, {Rees}  \& {Loeb}}{{Haiman}
  et~al.}{1997}]{1997ApJ...476..458H}
{Haiman} Z.,  {Rees} M.~J.,   {Loeb} A.,  1997, \mn@doi [\apj]
  {10.1086/303647}, \href
  {https://ui.adsabs.harvard.edu/abs/1997ApJ...476..458H} {476, 458}

\bibitem[\protect\citeauthoryear{{Hassan}, {Dav{\'e}}, {Finlator}  \&
  {Santos}}{{Hassan} et~al.}{2017}]{2017MNRAS.468..122H}
{Hassan} S.,  {Dav{\'e}} R.,  {Finlator} K.,   {Santos} M.~G.,  2017, \mn@doi
  [\mnras] {10.1093/mnras/stx420}, \href
  {https://ui.adsabs.harvard.edu/abs/2017MNRAS.468..122H} {468, 122}

\bibitem[\protect\citeauthoryear{{Hazra} \& {Smoot}}{{Hazra} \&
  {Smoot}}{2017}]{2017JCAP...11..028H}
{Hazra} D.~K.,  {Smoot} G.~F.,  2017, \mn@doi [\jcap]
  {10.1088/1475-7516/2017/11/028}, \href
  {https://ui.adsabs.harvard.edu/abs/2017JCAP...11..028H} {2017, 028}

\bibitem[\protect\citeauthoryear{{Hazra}, {Paoletti}, {Finelli}  \&
  {Smoot}}{{Hazra} et~al.}{2020}]{2020PhRvL.125g1301H}
{Hazra} D.~K.,  {Paoletti} D.,  {Finelli} F.,   {Smoot} G.~F.,  2020, \mn@doi
  [\prl] {10.1103/PhysRevLett.125.071301}, \href
  {https://ui.adsabs.harvard.edu/abs/2020PhRvL.125g1301H} {125, 071301}

\bibitem[\protect\citeauthoryear{{Hills}, {Kulkarni}, {Meerburg}  \&
  {Puchwein}}{{Hills} et~al.}{2018}]{2018Natur.564E..32H}
{Hills} R.,  {Kulkarni} G.,  {Meerburg} P.~D.,   {Puchwein} E.,  2018, \mn@doi
  [\nat] {10.1038/s41586-018-0796-5}, \href
  {https://ui.adsabs.harvard.edu/abs/2018Natur.564E..32H} {564, E32}

\bibitem[\protect\citeauthoryear{{Hu}}{{Hu}}{2000}]{2000ApJ...529...12H}
{Hu} W.,  2000, \mn@doi [\apj] {10.1086/308279}, \href
  {https://ui.adsabs.harvard.edu/abs/2000ApJ...529...12H} {529, 12}

\bibitem[\protect\citeauthoryear{{Hu} \& {Dodelson}}{{Hu} \&
  {Dodelson}}{2002}]{2002ARA&A..40..171H}
{Hu} W.,  {Dodelson} S.,  2002, \mn@doi [\araa]
  {10.1146/annurev.astro.40.060401.093926}, \href
  {https://ui.adsabs.harvard.edu/abs/2002ARA&A..40..171H} {40, 171}

\bibitem[\protect\citeauthoryear{{Hu} \& {Holder}}{{Hu} \&
  {Holder}}{2003}]{2003PhRvD..68b3001H}
{Hu} W.,  {Holder} G.~P.,  2003, \mn@doi [\prd] {10.1103/PhysRevD.68.023001},
  \href {https://ui.adsabs.harvard.edu/abs/2003PhRvD..68b3001H} {68, 023001}

\bibitem[\protect\citeauthoryear{{Hutter}, {Dayal}, {Yepes}, {Gottl{\"o}ber},
  {Legrand}  \& {Ucci}}{{Hutter} et~al.}{2020}]{2020arXiv200408401H}
{Hutter} A.,  {Dayal} P.,  {Yepes} G.,  {Gottl{\"o}ber} S.,  {Legrand} L.,
  {Ucci} G.,  2020, arXiv e-prints, \href
  {https://ui.adsabs.harvard.edu/abs/2020arXiv200408401H} {p. arXiv:2004.08401}

\bibitem[\protect\citeauthoryear{{Ishigaki}, {Kawamata}, {Ouchi}, {Oguri},
  {Shimasaku}  \& {Ono}}{{Ishigaki} et~al.}{2018}]{2018ApJ...854...73I}
{Ishigaki} M.,  {Kawamata} R.,  {Ouchi} M.,  {Oguri} M.,  {Shimasaku} K.,
  {Ono} Y.,  2018, \mn@doi [\apj] {10.3847/1538-4357/aaa544}, \href
  {https://ui.adsabs.harvard.edu/abs/2018ApJ...854...73I} {854, 73}

\bibitem[\protect\citeauthoryear{{Jaacks}, {Thompson}, {Finkelstein}  \&
  {Bromm}}{{Jaacks} et~al.}{2018}]{jaacks2018}
{Jaacks} J.,  {Thompson} R.,  {Finkelstein} S.~L.,   {Bromm} V.,  2018, \mn@doi
  [\mnras] {10.1093/mnras/sty062}, \href
  {https://ui.adsabs.harvard.edu/abs/2018MNRAS.475.4396J} {475, 4396}

\bibitem[\protect\citeauthoryear{{Katz}, {Kimm}, {Haehnelt}, {Sijacki},
  {Rosdahl}  \& {Blaizot}}{{Katz} et~al.}{2019}]{2019MNRAS.483.1029K}
{Katz} H.,  {Kimm} T.,  {Haehnelt} M.~G.,  {Sijacki} D.,  {Rosdahl} J.,
  {Blaizot} J.,  2019, \mn@doi [\mnras] {10.1093/mnras/sty3154}, \href
  {https://ui.adsabs.harvard.edu/abs/2019MNRAS.483.1029K} {483, 1029}

\bibitem[\protect\citeauthoryear{{Kern}, {Liu}, {Parsons}, {Mesinger}  \&
  {Greig}}{{Kern} et~al.}{2017}]{2017ApJ...848...23K}
{Kern} N.~S.,  {Liu} A.,  {Parsons} A.~R.,  {Mesinger} A.,   {Greig} B.,  2017,
  \mn@doi [\apj] {10.3847/1538-4357/aa8bb4}, \href
  {https://ui.adsabs.harvard.edu/abs/2017ApJ...848...23K} {848, 23}

\bibitem[\protect\citeauthoryear{{Kimm}, {Katz}, {Haehnelt}, {Rosdahl},
  {Devriendt}  \& {Slyz}}{{Kimm} et~al.}{2017}]{2017MNRAS.466.4826K}
{Kimm} T.,  {Katz} H.,  {Haehnelt} M.,  {Rosdahl} J.,  {Devriendt} J.,   {Slyz}
  A.,  2017, \mn@doi [\mnras] {10.1093/mnras/stx052}, \href
  {https://ui.adsabs.harvard.edu/abs/2017MNRAS.466.4826K} {466, 4826}

\bibitem[\protect\citeauthoryear{{Kimm}, {Blaizot}, {Garel}, {Michel-Dansac},
  {Katz}, {Rosdahl}, {Verhamme}  \& {Haehnelt}}{{Kimm}
  et~al.}{2019}]{2019MNRAS.486.2215K}
{Kimm} T.,  {Blaizot} J.,  {Garel} T.,  {Michel-Dansac} L.,  {Katz} H.,
  {Rosdahl} J.,  {Verhamme} A.,   {Haehnelt} M.,  2019, \mn@doi [\mnras]
  {10.1093/mnras/stz989}, \href
  {https://ui.adsabs.harvard.edu/abs/2019MNRAS.486.2215K} {486, 2215}

\bibitem[\protect\citeauthoryear{{Kulkarni}, {Keating}, {Haehnelt}, {Bosman},
  {Puchwein}, {Chardin}  \& {Aubert}}{{Kulkarni}
  et~al.}{2019a}]{2019MNRAS.485L..24K}
{Kulkarni} G.,  {Keating} L.~C.,  {Haehnelt} M.~G.,  {Bosman} S. E.~I.,
  {Puchwein} E.,  {Chardin} J.,   {Aubert} D.,  2019a, \mn@doi [\mnras]
  {10.1093/mnrasl/slz025}, \href
  {https://ui.adsabs.harvard.edu/abs/2019MNRAS.485L..24K} {485, L24}

\bibitem[\protect\citeauthoryear{{Kulkarni}, {Worseck}  \&
  {Hennawi}}{{Kulkarni} et~al.}{2019b}]{2019MNRAS.488.1035K}
{Kulkarni} G.,  {Worseck} G.,   {Hennawi} J.~F.,  2019b, \mn@doi [\mnras]
  {10.1093/mnras/stz1493}, \href
  {https://ui.adsabs.harvard.edu/abs/2019MNRAS.488.1035K} {488, 1035}

\bibitem[\protect\citeauthoryear{{Lesgourgues}}{{Lesgourgues}}{2011}]{2011arXiv1104.2932L}
{Lesgourgues} J.,  2011, arXiv e-prints, \href
  {https://ui.adsabs.harvard.edu/abs/2011arXiv1104.2932L} {p. arXiv:1104.2932}

\bibitem[\protect\citeauthoryear{{Lewis}}{{Lewis}}{2008}]{2008PhRvD..78b3002L}
{Lewis} A.,  2008, \mn@doi [\prd] {10.1103/PhysRevD.78.023002}, \href
  {https://ui.adsabs.harvard.edu/abs/2008PhRvD..78b3002L} {78, 023002}

\bibitem[\protect\citeauthoryear{Lewis}{Lewis}{2013}]{Lewis:2013hha}
Lewis A.,  2013, \mn@doi [\prd] {10.1103/PhysRevD.87.103529}, 87, 103529

\bibitem[\protect\citeauthoryear{{Liu} \& {Parsons}}{{Liu} \&
  {Parsons}}{2016}]{2016MNRAS.457.1864L}
{Liu} A.,  {Parsons} A.~R.,  2016, \mn@doi [\mnras] {10.1093/mnras/stw071},
  \href {https://ui.adsabs.harvard.edu/abs/2016MNRAS.457.1864L} {457, 1864}

\bibitem[\protect\citeauthoryear{{Liu}, {Pritchard}, {Allison}, {Parsons},
  {Seljak}  \& {Sherwin}}{{Liu} et~al.}{2016}]{2016PhRvD..93d3013L}
{Liu} A.,  {Pritchard} J.~R.,  {Allison} R.,  {Parsons} A.~R.,  {Seljak} U.,
  {Sherwin} B.~D.,  2016, \mn@doi [\prd] {10.1103/PhysRevD.93.043013}, \href
  {https://ui.adsabs.harvard.edu/abs/2016PhRvD..93d3013L} {93, 043013}

\bibitem[\protect\citeauthoryear{{Livermore}, {Finkelstein}  \&
  {Lotz}}{{Livermore} et~al.}{2017}]{2017ApJ...835..113L}
{Livermore} R.~C.,  {Finkelstein} S.~L.,   {Lotz} J.~M.,  2017, \mn@doi [\apj]
  {10.3847/1538-4357/835/2/113}, \href
  {https://ui.adsabs.harvard.edu/abs/2017ApJ...835..113L} {835, 113}

\bibitem[\protect\citeauthoryear{{Loeb} \& {Barkana}}{{Loeb} \&
  {Barkana}}{2001}]{2001ARA&A..39...19L}
{Loeb} A.,  {Barkana} R.,  2001, \mn@doi [\araa]
  {10.1146/annurev.astro.39.1.19}, \href
  {https://ui.adsabs.harvard.edu/abs/2001ARA&A..39...19L} {39, 19}

\bibitem[\protect\citeauthoryear{{Maio}, {Ciardi}, {Dolag}, {Tornatore}  \&
  {Khochfar}}{{Maio} et~al.}{2010}]{maio2010}
{Maio} U.,  {Ciardi} B.,  {Dolag} K.,  {Tornatore} L.,   {Khochfar} S.,  2010,
  \mn@doi [\mnras] {10.1111/j.1365-2966.2010.17003.x}, \href
  {http://adsabs.harvard.edu/abs/2010MNRAS.407.1003M} {407, 1003}

\bibitem[\protect\citeauthoryear{{Maio}, {Khochfar}, {Johnson}  \&
  {Ciardi}}{{Maio} et~al.}{2011}]{maio2011}
{Maio} U.,  {Khochfar} S.,  {Johnson} J.~L.,   {Ciardi} B.,  2011, \mnras,
  \href {http://adsabs.harvard.edu/abs/2011MNRAS.414.1145M} {414, 1145}

\bibitem[\protect\citeauthoryear{{McGreer}, {Mesinger}  \&
  {D'Odorico}}{{McGreer} et~al.}{2015}]{2015MNRAS.447..499M}
{McGreer} I.~D.,  {Mesinger} A.,   {D'Odorico} V.,  2015, \mn@doi [\mnras]
  {10.1093/mnras/stu2449}, \href
  {https://ui.adsabs.harvard.edu/abs/2015MNRAS.447..499M} {447, 499}

\bibitem[\protect\citeauthoryear{{McQuinn}, {Furlanetto}, {Hernquist}, {Zahn}
  \& {Zaldarriaga}}{{McQuinn} et~al.}{2005}]{2005ApJ...630..643M}
{McQuinn} M.,  {Furlanetto} S.~R.,  {Hernquist} L.,  {Zahn} O.,   {Zaldarriaga}
  M.,  2005, \mn@doi [\apj] {10.1086/432049}, \href
  {https://ui.adsabs.harvard.edu/abs/2005ApJ...630..643M} {630, 643}

\bibitem[\protect\citeauthoryear{{Mebane}, {Mirocha}  \& {Furlanetto}}{{Mebane}
  et~al.}{2020}]{2020MNRAS.493.1217M}
{Mebane} R.~H.,  {Mirocha} J.,   {Furlanetto} S.~R.,  2020, \mn@doi [\mnras]
  {10.1093/mnras/staa280}, \href
  {https://ui.adsabs.harvard.edu/abs/2020MNRAS.493.1217M} {493, 1217}

\bibitem[\protect\citeauthoryear{{Mesinger}, {Greig}  \& {Sobacchi}}{{Mesinger}
  et~al.}{2016}]{2016MNRAS.459.2342M}
{Mesinger} A.,  {Greig} B.,   {Sobacchi} E.,  2016, \mn@doi [\mnras]
  {10.1093/mnras/stw831}, \href
  {https://ui.adsabs.harvard.edu/abs/2016MNRAS.459.2342M} {459, 2342}

\bibitem[\protect\citeauthoryear{{Mineo}, {Gilfanov}  \& {Sunyaev}}{{Mineo}
  et~al.}{2012}]{2012MNRAS.419.2095M}
{Mineo} S.,  {Gilfanov} M.,   {Sunyaev} R.,  2012, \mn@doi [\mnras]
  {10.1111/j.1365-2966.2011.19862.x}, \href
  {https://ui.adsabs.harvard.edu/abs/2012MNRAS.419.2095M} {419, 2095}

\bibitem[\protect\citeauthoryear{{Miralda-Escud{\'e}}}{{Miralda-Escud{\'e}}}{2003}]{2003ApJ...597...66M}
{Miralda-Escud{\'e}} J.,  2003, \mn@doi [\apj] {10.1086/378286}, \href
  {https://ui.adsabs.harvard.edu/abs/2003ApJ...597...66M} {597, 66}

\bibitem[\protect\citeauthoryear{{Miranda}, {Lidz}, {Heinrich}  \&
  {Hu}}{{Miranda} et~al.}{2017}]{2017MNRAS.467.4050M}
{Miranda} V.,  {Lidz} A.,  {Heinrich} C.~H.,   {Hu} W.,  2017, \mn@doi [\mnras]
  {10.1093/mnras/stx306}, \href
  {https://ui.adsabs.harvard.edu/abs/2017MNRAS.467.4050M} {467, 4050}

\bibitem[\protect\citeauthoryear{{Mirocha} \& {Furlanetto}}{{Mirocha} \&
  {Furlanetto}}{2019}]{2019MNRAS.483.1980M}
{Mirocha} J.,  {Furlanetto} S.~R.,  2019, \mn@doi [\mnras]
  {10.1093/mnras/sty3260}, \href
  {https://ui.adsabs.harvard.edu/abs/2019MNRAS.483.1980M} {483, 1980}

\bibitem[\protect\citeauthoryear{{Mitra}, {Choudhury}  \& {Ferrara}}{{Mitra}
  et~al.}{2011}]{2011MNRAS.413.1569M}
{Mitra} S.,  {Choudhury} T.~R.,   {Ferrara} A.,  2011, \mn@doi [\mnras]
  {10.1111/j.1365-2966.2011.18234.x}, \href
  {https://ui.adsabs.harvard.edu/abs/2011MNRAS.413.1569M} {413, 1569}

\bibitem[\protect\citeauthoryear{{Mitra}, {Choudhury}  \& {Ferrara}}{{Mitra}
  et~al.}{2012}]{2012MNRAS.419.1480M}
{Mitra} S.,  {Choudhury} T.~R.,   {Ferrara} A.,  2012, \mn@doi [\mnras]
  {10.1111/j.1365-2966.2011.19804.x}, \href
  {https://ui.adsabs.harvard.edu/abs/2012MNRAS.419.1480M} {419, 1480}

\bibitem[\protect\citeauthoryear{{Mitra}, {Ferrara}  \& {Choudhury}}{{Mitra}
  et~al.}{2013}]{2013MNRAS.428L...1M}
{Mitra} S.,  {Ferrara} A.,   {Choudhury} T.~R.,  2013, \mn@doi [\mnras]
  {10.1093/mnrasl/sls001}, \href
  {https://ui.adsabs.harvard.edu/abs/2013MNRAS.428L...1M} {428, L1}

\bibitem[\protect\citeauthoryear{{Mitra}, {Choudhury}  \& {Ferrara}}{{Mitra}
  et~al.}{2015}]{2015MNRAS.454L..76M}
{Mitra} S.,  {Choudhury} T.~R.,   {Ferrara} A.,  2015, \mn@doi [\mnras]
  {10.1093/mnrasl/slv134}, \href
  {https://ui.adsabs.harvard.edu/abs/2015MNRAS.454L..76M} {454, L76}

\bibitem[\protect\citeauthoryear{{Mitra}, {Choudhury}  \& {Ferrara}}{{Mitra}
  et~al.}{2018a}]{2018MNRAS.473.1416M}
{Mitra} S.,  {Choudhury} T.~R.,   {Ferrara} A.,  2018a, \mn@doi [\mnras]
  {10.1093/mnras/stx2443}, \href
  {https://ui.adsabs.harvard.edu/abs/2018MNRAS.473.1416M} {473, 1416}

\bibitem[\protect\citeauthoryear{{Mitra}, {Choudhury}  \& {Ratra}}{{Mitra}
  et~al.}{2018b}]{2018MNRAS.479.4566M}
{Mitra} S.,  {Choudhury} T.~R.,   {Ratra} B.,  2018b, \mn@doi [\mnras]
  {10.1093/mnras/sty1835}, \href
  {https://ui.adsabs.harvard.edu/abs/2018MNRAS.479.4566M} {479, 4566}

\bibitem[\protect\citeauthoryear{{Mitra}, {Park}, {Choudhury}  \&
  {Ratra}}{{Mitra} et~al.}{2019}]{2019MNRAS.487.5118M}
{Mitra} S.,  {Park} C.-G.,  {Choudhury} T.~R.,   {Ratra} B.,  2019, \mn@doi
  [\mnras] {10.1093/mnras/stz1560}, \href
  {https://ui.adsabs.harvard.edu/abs/2019MNRAS.487.5118M} {487, 5118}

\bibitem[\protect\citeauthoryear{{Mittal} \& {Kulkarni}}{{Mittal} \&
  {Kulkarni}}{2021}]{2021MNRAS.503.4264M}
{Mittal} S.,  {Kulkarni} G.,  2021, \mn@doi [\mnras] {10.1093/mnras/staa3811},
  \href {https://ui.adsabs.harvard.edu/abs/2021MNRAS.503.4264M} {503, 4264}

\bibitem[\protect\citeauthoryear{{Mortonson} \& {Hu}}{{Mortonson} \&
  {Hu}}{2008}]{2008ApJ...672..737M}
{Mortonson} M.~J.,  {Hu} W.,  2008, \mn@doi [\apj] {10.1086/523958}, \href
  {https://ui.adsabs.harvard.edu/abs/2008ApJ...672..737M} {672, 737}

\bibitem[\protect\citeauthoryear{{Nhan}, {Bordenave}, {Bradley}, {Burns},
  {Tauscher}, {Rapetti}  \& {Klima}}{{Nhan} et~al.}{2018}]{Nahn_2018}
{Nhan} B.~D.,  {Bordenave} D.~D.,  {Bradley} R.~F.,  {Burns} J.~O.,  {Tauscher}
  K.,  {Rapetti} D.,   {Klima} P.~J.,  2018, arXiv e-prints, \href
  {https://ui.adsabs.harvard.edu/abs/2018arXiv181104917N} {p. arXiv:1811.04917}

\bibitem[\protect\citeauthoryear{{O'Meara}, {Prochaska}, {Worseck}, {Chen}  \&
  {Madau}}{{O'Meara} et~al.}{2013}]{2013ApJ...765..137O}
{O'Meara} J.~M.,  {Prochaska} J.~X.,  {Worseck} G.,  {Chen} H.-W.,   {Madau}
  P.,  2013, \mn@doi [\apj] {10.1088/0004-637X/765/2/137}, \href
  {https://ui.adsabs.harvard.edu/abs/2013ApJ...765..137O} {765, 137}

\bibitem[\protect\citeauthoryear{{Ocvirk} et~al.,}{{Ocvirk}
  et~al.}{2016}]{2016MNRAS.463.1462O}
{Ocvirk} P.,  et~al., 2016, \mn@doi [\mnras] {10.1093/mnras/stw2036}, \href
  {https://ui.adsabs.harvard.edu/abs/2016MNRAS.463.1462O} {463, 1462}

\bibitem[\protect\citeauthoryear{{Ocvirk}, {Aubert}, {Chardin}, {Deparis}  \&
  {Lewis}}{{Ocvirk} et~al.}{2019}]{2019A&A...626A..77O}
{Ocvirk} P.,  {Aubert} D.,  {Chardin} J.,  {Deparis} N.,   {Lewis} J.,  2019,
  \mn@doi [\aap] {10.1051/0004-6361/201832923}, \href
  {https://ui.adsabs.harvard.edu/abs/2019A&A...626A..77O} {626, A77}

\bibitem[\protect\citeauthoryear{{Oesch} et~al.,}{{Oesch}
  et~al.}{2014}]{2014ApJ...786..108O}
{Oesch} P.~A.,  et~al., 2014, \mn@doi [\apj] {10.1088/0004-637X/786/2/108},
  \href {https://ui.adsabs.harvard.edu/abs/2014ApJ...786..108O} {786, 108}

\bibitem[\protect\citeauthoryear{{Oesch}, {Bouwens}, {Illingworth}, {Labb{\'e}}
   \& {Stefanon}}{{Oesch} et~al.}{2018}]{2018ApJ...855..105O}
{Oesch} P.~A.,  {Bouwens} R.~J.,  {Illingworth} G.~D.,  {Labb{\'e}} I.,
  {Stefanon} M.,  2018, \mn@doi [\apj] {10.3847/1538-4357/aab03f}, \href
  {https://ui.adsabs.harvard.edu/abs/2018ApJ...855..105O} {855, 105}

\bibitem[\protect\citeauthoryear{{Omukai} \& {Palla}}{{Omukai} \&
  {Palla}}{2001}]{2001ApJ...561L..55O}
{Omukai} K.,  {Palla} F.,  2001, \mn@doi [\apjl] {10.1086/324410}, \href
  {https://ui.adsabs.harvard.edu/abs/2001ApJ...561L..55O} {561, L55}

\bibitem[\protect\citeauthoryear{{Paardekooper}, {Khochfar}  \& {Dalla
  Vecchia}}{{Paardekooper} et~al.}{2015}]{2015MNRAS.451.2544P}
{Paardekooper} J.-P.,  {Khochfar} S.,   {Dalla Vecchia} C.,  2015, \mn@doi
  [\mnras] {10.1093/mnras/stv1114}, \href
  {https://ui.adsabs.harvard.edu/abs/2015MNRAS.451.2544P} {451, 2544}

\bibitem[\protect\citeauthoryear{{Pandey}, {Choudhury}, {Sethi}  \&
  {Ferrara}}{{Pandey} et~al.}{2015}]{2015MNRAS.451.1692P}
{Pandey} K.~L.,  {Choudhury} T.~R.,  {Sethi} S.~K.,   {Ferrara} A.,  2015,
  \mn@doi [\mnras] {10.1093/mnras/stv1055}, \href
  {https://ui.adsabs.harvard.edu/abs/2015MNRAS.451.1692P} {451, 1692}

\bibitem[\protect\citeauthoryear{{Pandolfi}, {Ferrara}, {Choudhury},
  {Melchiorri}  \& {Mitra}}{{Pandolfi} et~al.}{2011}]{2011PhRvD..84l3522P}
{Pandolfi} S.,  {Ferrara} A.,  {Choudhury} T.~R.,  {Melchiorri} A.,   {Mitra}
  S.,  2011, \mn@doi [\prd] {10.1103/PhysRevD.84.123522}, \href
  {https://ui.adsabs.harvard.edu/abs/2011PhRvD..84l3522P} {84, 123522}

\bibitem[\protect\citeauthoryear{{Park}, {Mesinger}, {Greig}  \&
  {Gillet}}{{Park} et~al.}{2019}]{2019MNRAS.484..933P}
{Park} J.,  {Mesinger} A.,  {Greig} B.,   {Gillet} N.,  2019, \mn@doi [\mnras]
  {10.1093/mnras/stz032}, \href
  {https://ui.adsabs.harvard.edu/abs/2019MNRAS.484..933P} {484, 933}

\bibitem[\protect\citeauthoryear{{Park}, {Gillet}, {Mesinger}  \&
  {Greig}}{{Park} et~al.}{2020}]{2020MNRAS.491.3891P}
{Park} J.,  {Gillet} N.,  {Mesinger} A.,   {Greig} B.,  2020, \mn@doi [\mnras]
  {10.1093/mnras/stz3278}, \href
  {https://ui.adsabs.harvard.edu/abs/2020MNRAS.491.3891P} {491, 3891}

\bibitem[\protect\citeauthoryear{{Patra}, {Subrahmanyan}, {Raghunathan}  \&
  {Udaya Shankar}}{{Patra} et~al.}{2013}]{Patra_2013}
{Patra} N.,  {Subrahmanyan} R.,  {Raghunathan} A.,   {Udaya Shankar} N.,  2013,
  \mn@doi [Experimental Astronomy] {10.1007/s10686-013-9336-3}, \href
  {http://adsabs.harvard.edu/abs/2013ExA....36..319P} {36, 319}

\bibitem[\protect\citeauthoryear{{Planck Collaboration} et~al.,}{{Planck
  Collaboration} et~al.}{2014}]{2014A&A...571A..15P}
{Planck Collaboration} et~al., 2014, \mn@doi [\aap]
  {10.1051/0004-6361/201321573}, \href
  {https://ui.adsabs.harvard.edu/abs/2014A&A...571A..15P} {571, A15}

\bibitem[\protect\citeauthoryear{{Planck Collaboration} et~al.,}{{Planck
  Collaboration} et~al.}{2016}]{2016A&A...594A..13P}
{Planck Collaboration} et~al., 2016, \mn@doi [\aap]
  {10.1051/0004-6361/201525830}, \href
  {https://ui.adsabs.harvard.edu/abs/2016A&A...594A..13P} {594, A13}

\bibitem[\protect\citeauthoryear{{Planck Collaboration} et~al.,}{{Planck
  Collaboration} et~al.}{2020}]{2020A&A...641A...6P}
{Planck Collaboration} et~al., 2020, \mn@doi [\aap]
  {10.1051/0004-6361/201833910}, \href
  {https://ui.adsabs.harvard.edu/abs/2020A&A...641A...6P} {641, A6}

\bibitem[\protect\citeauthoryear{{Pospelov}, {Pradler}, {Ruderman}  \&
  {Urbano}}{{Pospelov} et~al.}{2018}]{2018PhRvL.121c1103P}
{Pospelov} M.,  {Pradler} J.,  {Ruderman} J.~T.,   {Urbano} A.,  2018, \mn@doi
  [\prl] {10.1103/PhysRevLett.121.031103}, \href
  {https://ui.adsabs.harvard.edu/abs/2018PhRvL.121c1103P} {121, 031103}

\bibitem[\protect\citeauthoryear{{Press} \& {Schechter}}{{Press} \&
  {Schechter}}{1974}]{1974ApJ...187..425P}
{Press} W.~H.,  {Schechter} P.,  1974, \mn@doi [\apj] {10.1086/152650}, \href
  {https://ui.adsabs.harvard.edu/abs/1974ApJ...187..425P} {187, 425}

\bibitem[\protect\citeauthoryear{{Prochaska}, {O'Meara}  \&
  {Worseck}}{{Prochaska} et~al.}{2010}]{2010ApJ...718..392P}
{Prochaska} J.~X.,  {O'Meara} J.~M.,   {Worseck} G.,  2010, \mn@doi [\apj]
  {10.1088/0004-637X/718/1/392}, \href
  {https://ui.adsabs.harvard.edu/abs/2010ApJ...718..392P} {718, 392}

\bibitem[\protect\citeauthoryear{{Qin} et~al.,}{{Qin}
  et~al.}{2017}]{2017MNRAS.472.2009Q}
{Qin} Y.,  et~al., 2017, \mn@doi [\mnras] {10.1093/mnras/stx1909}, \href
  {https://ui.adsabs.harvard.edu/abs/2017MNRAS.472.2009Q} {472, 2009}

\bibitem[\protect\citeauthoryear{{Qin}, {Poulin}, {Mesinger}, {Greig}, {Murray}
   \& {Park}}{{Qin} et~al.}{2020}]{2020MNRAS.tmp.2712Q}
{Qin} Y.,  {Poulin} V.,  {Mesinger} A.,  {Greig} B.,  {Murray} S.,   {Park} J.,
   2020, \mn@doi [\mnras] {10.1093/mnras/staa2797}, \href
  {https://ui.adsabs.harvard.edu/abs/2020MNRAS.tmp.2712Q} {}

\bibitem[\protect\citeauthoryear{{Reis}, {Fialkov}  \& {Barkana}}{{Reis}
  et~al.}{2020}]{2020arXiv200804315R}
{Reis} I.,  {Fialkov} A.,   {Barkana} R.,  2020, arXiv e-prints, \href
  {https://ui.adsabs.harvard.edu/abs/2020arXiv200804315R} {p. arXiv:2008.04315}

\bibitem[\protect\citeauthoryear{{Ribaudo}, {Lehner}  \& {Howk}}{{Ribaudo}
  et~al.}{2011}]{2011ApJ...736...42R}
{Ribaudo} J.,  {Lehner} N.,   {Howk} J.~C.,  2011, \mn@doi [\apj]
  {10.1088/0004-637X/736/1/42}, \href
  {https://ui.adsabs.harvard.edu/abs/2011ApJ...736...42R} {736, 42}

\bibitem[\protect\citeauthoryear{{Sarmento}, {Scannapieco}  \&
  {C{\^o}t{\'e}}}{{Sarmento} et~al.}{2019}]{sarmento2019}
{Sarmento} R.,  {Scannapieco} E.,   {C{\^o}t{\'e}} B.,  2019, \mn@doi [\apj]
  {10.3847/1538-4357/aafa1a}, \href
  {https://ui.adsabs.harvard.edu/abs/2019ApJ...871..206S} {871, 206}

\bibitem[\protect\citeauthoryear{{Schaerer}}{{Schaerer}}{2002}]{2002A&A...382...28S}
{Schaerer} D.,  2002, \mn@doi [\aap] {10.1051/0004-6361:20011619}, \href
  {https://ui.adsabs.harvard.edu/abs/2002A&A...382...28S} {382, 28}

\bibitem[\protect\citeauthoryear{{Schauer}, {Liu}  \& {Bromm}}{{Schauer}
  et~al.}{2019}]{2019ApJ...877L...5S}
{Schauer} A. T.~P.,  {Liu} B.,   {Bromm} V.,  2019, \mn@doi [\apjl]
  {10.3847/2041-8213/ab1e51}, \href
  {https://ui.adsabs.harvard.edu/abs/2019ApJ...877L...5S} {877, L5}

\bibitem[\protect\citeauthoryear{{Schirber} \& {Bullock}}{{Schirber} \&
  {Bullock}}{2003}]{2003ApJ...584..110S}
{Schirber} M.,  {Bullock} J.~S.,  2003, \mn@doi [\apj] {10.1086/345662}, \href
  {https://ui.adsabs.harvard.edu/abs/2003ApJ...584..110S} {584, 110}

\bibitem[\protect\citeauthoryear{{Schmit} \& {Pritchard}}{{Schmit} \&
  {Pritchard}}{2018}]{2018MNRAS.475.1213S}
{Schmit} C.~J.,  {Pritchard} J.~R.,  2018, \mn@doi [\mnras]
  {10.1093/mnras/stx3292}, \href
  {https://ui.adsabs.harvard.edu/abs/2018MNRAS.475.1213S} {475, 1213}

\bibitem[\protect\citeauthoryear{{Seiffert} et~al.,}{{Seiffert}
  et~al.}{2011}]{2011ApJ...734....6S}
{Seiffert} M.,  et~al., 2011, \mn@doi [\apj] {10.1088/0004-637X/734/1/6}, \href
  {https://ui.adsabs.harvard.edu/abs/2011ApJ...734....6S} {734, 6}

\bibitem[\protect\citeauthoryear{{Shang}, {Bryan}  \& {Haiman}}{{Shang}
  et~al.}{2010}]{2010MNRAS.402.1249S}
{Shang} C.,  {Bryan} G.~L.,   {Haiman} Z.,  2010, \mn@doi [\mnras]
  {10.1111/j.1365-2966.2009.15960.x}, \href
  {https://ui.adsabs.harvard.edu/abs/2010MNRAS.402.1249S} {402, 1249}

\bibitem[\protect\citeauthoryear{{Sharma}}{{Sharma}}{2018}]{2018MNRAS.481L...6S}
{Sharma} P.,  2018, \mn@doi [\mnras] {10.1093/mnrasl/sly147}, \href
  {https://ui.adsabs.harvard.edu/abs/2018MNRAS.481L...6S} {481, L6}

\bibitem[\protect\citeauthoryear{{Sheth} \& {Tormen}}{{Sheth} \&
  {Tormen}}{1999}]{1999MNRAS.308..119S}
{Sheth} R.~K.,  {Tormen} G.,  1999, \mn@doi [\mnras]
  {10.1046/j.1365-8711.1999.02692.x}, \href
  {https://ui.adsabs.harvard.edu/abs/1999MNRAS.308..119S} {308, 119}

\bibitem[\protect\citeauthoryear{{Sheth}, {Mo}  \& {Tormen}}{{Sheth}
  et~al.}{2001}]{2001MNRAS.323....1S}
{Sheth} R.~K.,  {Mo} H.~J.,   {Tormen} G.,  2001, \mn@doi [\mnras]
  {10.1046/j.1365-8711.2001.04006.x}, \href
  {https://ui.adsabs.harvard.edu/abs/2001MNRAS.323....1S} {323, 1}

\bibitem[\protect\citeauthoryear{{Sims} \& {Pober}}{{Sims} \&
  {Pober}}{2020}]{2020MNRAS.492...22S}
{Sims} P.~H.,  {Pober} J.~C.,  2020, \mn@doi [\mnras] {10.1093/mnras/stz3388},
  \href {https://ui.adsabs.harvard.edu/abs/2020MNRAS.492...22S} {492, 22}

\bibitem[\protect\citeauthoryear{{Singh} \& {Subrahmanyan}}{{Singh} \&
  {Subrahmanyan}}{2019}]{2019ApJ...880...26S}
{Singh} S.,  {Subrahmanyan} R.,  2019, \mn@doi [\apj]
  {10.3847/1538-4357/ab2879}, \href
  {https://ui.adsabs.harvard.edu/abs/2019ApJ...880...26S} {880, 26}

\bibitem[\protect\citeauthoryear{{Singh} et~al.,}{{Singh}
  et~al.}{2017}]{Singh_2017}
{Singh} S.,  et~al., 2017, \mn@doi [\apjl] {10.3847/2041-8213/aa831b}, \href
  {http://adsabs.harvard.edu/abs/2017ApJ...845L..12S} {845, L12}

\bibitem[\protect\citeauthoryear{{Slatyer} \& {Wu}}{{Slatyer} \&
  {Wu}}{2018}]{2018PhRvD..98b3013S}
{Slatyer} T.~R.,  {Wu} C.-L.,  2018, \mn@doi [\prd]
  {10.1103/PhysRevD.98.023013}, \href
  {https://ui.adsabs.harvard.edu/abs/2018PhRvD..98b3013S} {98, 023013}

\bibitem[\protect\citeauthoryear{{Sokolowski} et~al.,}{{Sokolowski}
  et~al.}{2015}]{Sokolowski_2015}
{Sokolowski} M.,  et~al., 2015, \mn@doi [\pasa] {10.1017/pasa.2015.3}, \href
  {http://adsabs.harvard.edu/abs/2015PASA...32....4S} {32, e004}

\bibitem[\protect\citeauthoryear{{Songaila} \& {Cowie}}{{Songaila} \&
  {Cowie}}{2010}]{2010ApJ...721.1448S}
{Songaila} A.,  {Cowie} L.~L.,  2010, \mn@doi [\apj]
  {10.1088/0004-637X/721/2/1448}, \href
  {https://ui.adsabs.harvard.edu/abs/2010ApJ...721.1448S} {721, 1448}

\bibitem[\protect\citeauthoryear{{Sunyaev} \& {Zeldovich}}{{Sunyaev} \&
  {Zeldovich}}{1980}]{1980MNRAS.190..413S}
{Sunyaev} R.~A.,  {Zeldovich} Y.~B.,  1980, \mn@doi [\mnras]
  {10.1093/mnras/190.3.413}, \href
  {https://ui.adsabs.harvard.edu/abs/1980MNRAS.190..413S} {190, 413}

\bibitem[\protect\citeauthoryear{{Telfer}, {Zheng}, {Kriss}  \&
  {Davidsen}}{{Telfer} et~al.}{2002}]{2002ApJ...565..773T}
{Telfer} R.~C.,  {Zheng} W.,  {Kriss} G.~A.,   {Davidsen} A.~F.,  2002, \mn@doi
  [\apj] {10.1086/324689}, \href
  {https://ui.adsabs.harvard.edu/abs/2002ApJ...565..773T} {565, 773}

\bibitem[\protect\citeauthoryear{{Valiante}, {Schneider}, {Volonteri}  \&
  {Omukai}}{{Valiante} et~al.}{2016}]{2016MNRAS.457.3356V}
{Valiante} R.,  {Schneider} R.,  {Volonteri} M.,   {Omukai} K.,  2016, \mn@doi
  [\mnras] {10.1093/mnras/stw225}, \href
  {https://ui.adsabs.harvard.edu/abs/2016MNRAS.457.3356V} {457, 3356}

\bibitem[\protect\citeauthoryear{{Visbal}, {Haiman}, {Terrazas}, {Bryan}  \&
  {Barkana}}{{Visbal} et~al.}{2014}]{2014MNRAS.445..107V}
{Visbal} E.,  {Haiman} Z.,  {Terrazas} B.,  {Bryan} G.~L.,   {Barkana} R.,
  2014, \mn@doi [\mnras] {10.1093/mnras/stu1710}, \href
  {https://ui.adsabs.harvard.edu/abs/2014MNRAS.445..107V} {445, 107}

\bibitem[\protect\citeauthoryear{{Vishniac}}{{Vishniac}}{1987}]{1987ApJ...322..597V}
{Vishniac} E.~T.,  1987, \mn@doi [\apj] {10.1086/165755}, \href
  {https://ui.adsabs.harvard.edu/abs/1987ApJ...322..597V} {322, 597}

\bibitem[\protect\citeauthoryear{{Voytek}, {Natarajan}, {J{\'a}uregui
  Garc{\'{\i}}a}, {Peterson}  \& {L{\'o}pez-Cruz}}{{Voytek}
  et~al.}{2014}]{Voytek_2014}
{Voytek} T.~C.,  {Natarajan} A.,  {J{\'a}uregui Garc{\'{\i}}a} J.~M.,
  {Peterson} J.~B.,   {L{\'o}pez-Cruz} O.,  2014, \mn@doi [\apjl]
  {10.1088/2041-8205/782/1/L9}, \href
  {http://adsabs.harvard.edu/abs/2014ApJ...782L...9V} {782, L9}

\bibitem[\protect\citeauthoryear{{Wise}, {Demchenko}, {Halicek}, {Norman},
  {Turk}, {Abel}  \& {Smith}}{{Wise} et~al.}{2014}]{2014MNRAS.442.2560W}
{Wise} J.~H.,  {Demchenko} V.~G.,  {Halicek} M.~T.,  {Norman} M.~L.,  {Turk}
  M.~J.,  {Abel} T.,   {Smith} B.~D.,  2014, \mn@doi [\mnras]
  {10.1093/mnras/stu979}, \href
  {https://ui.adsabs.harvard.edu/abs/2014MNRAS.442.2560W} {442, 2560}

\bibitem[\protect\citeauthoryear{{Wolcott-Green}, {Haiman}  \&
  {Bryan}}{{Wolcott-Green} et~al.}{2011}]{2011MNRAS.418..838W}
{Wolcott-Green} J.,  {Haiman} Z.,   {Bryan} G.~L.,  2011, \mn@doi [\mnras]
  {10.1111/j.1365-2966.2011.19538.x}, \href
  {https://ui.adsabs.harvard.edu/abs/2011MNRAS.418..838W} {418, 838}

\bibitem[\protect\citeauthoryear{{Wu}, {McQuinn}, {Kannan}, {D'Aloisio},
  {Bird}, {Marinacci}, {Dav{\'e}}  \& {Hernquist}}{{Wu}
  et~al.}{2019}]{2019MNRAS.490.3177W}
{Wu} X.,  {McQuinn} M.,  {Kannan} R.,  {D'Aloisio} A.,  {Bird} S.,  {Marinacci}
  F.,  {Dav{\'e}} R.,   {Hernquist} L.,  2019, \mn@doi [\mnras]
  {10.1093/mnras/stz2807}, \href
  {https://ui.adsabs.harvard.edu/abs/2019MNRAS.490.3177W} {490, 3177}

\bibitem[\protect\citeauthoryear{{Yajima}, {Choi}  \& {Nagamine}}{{Yajima}
  et~al.}{2012}]{2012MNRAS.427.2889Y}
{Yajima} H.,  {Choi} J.-H.,   {Nagamine} K.,  2012, \mn@doi [\mnras]
  {10.1111/j.1365-2966.2012.22131.x}, \href
  {https://ui.adsabs.harvard.edu/abs/2012MNRAS.427.2889Y} {427, 2889}

\bibitem[\protect\citeauthoryear{{Yung}, {Somerville}, {Popping},
  {Finkelstein}, {Ferguson}  \& {Dav{\'e}}}{{Yung}
  et~al.}{2019}]{2019MNRAS.490.2855Y}
{Yung} L.~Y.~A.,  {Somerville} R.~S.,  {Popping} G.,  {Finkelstein} S.~L.,
  {Ferguson} H.~C.,   {Dav{\'e}} R.,  2019, \mn@doi [\mnras]
  {10.1093/mnras/stz2755}, \href
  {https://ui.adsabs.harvard.edu/abs/2019MNRAS.490.2855Y} {490, 2855}

\makeatother
\end{thebibliography}



\appendix

\section{Comparison with the ``traditional'' constraints from CMB}

\begin{figure*}
    \centering
    \includegraphics[width=0.7\textwidth]{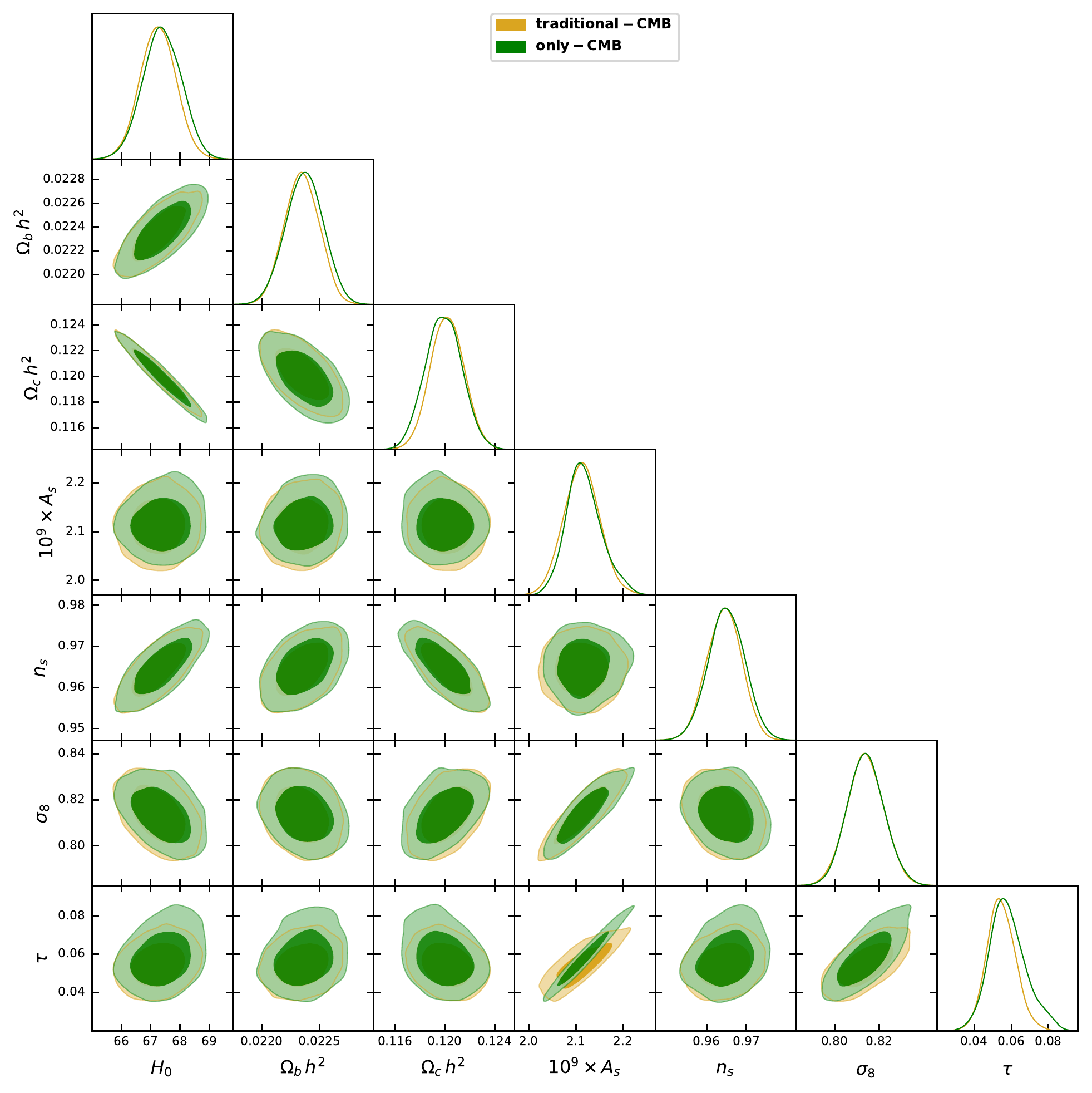}
    \caption{The marginalized posterior distributions of cosmological parameters obtained from \textbf{only-CMB} (green) and \textbf{traditional-CMB} (orange) analyses. We show the $68 \%$ and $95\%$ confidence contours in the two-dimensional plots. Note that $\sigma_8$ and $\tau$ are derived parameters while the others are free parameters in our model. }
    \label{fig:comparing_tanh_CF}
\end{figure*}

\begin{figure}
    \centering
    \includegraphics[width=0.5\textwidth]{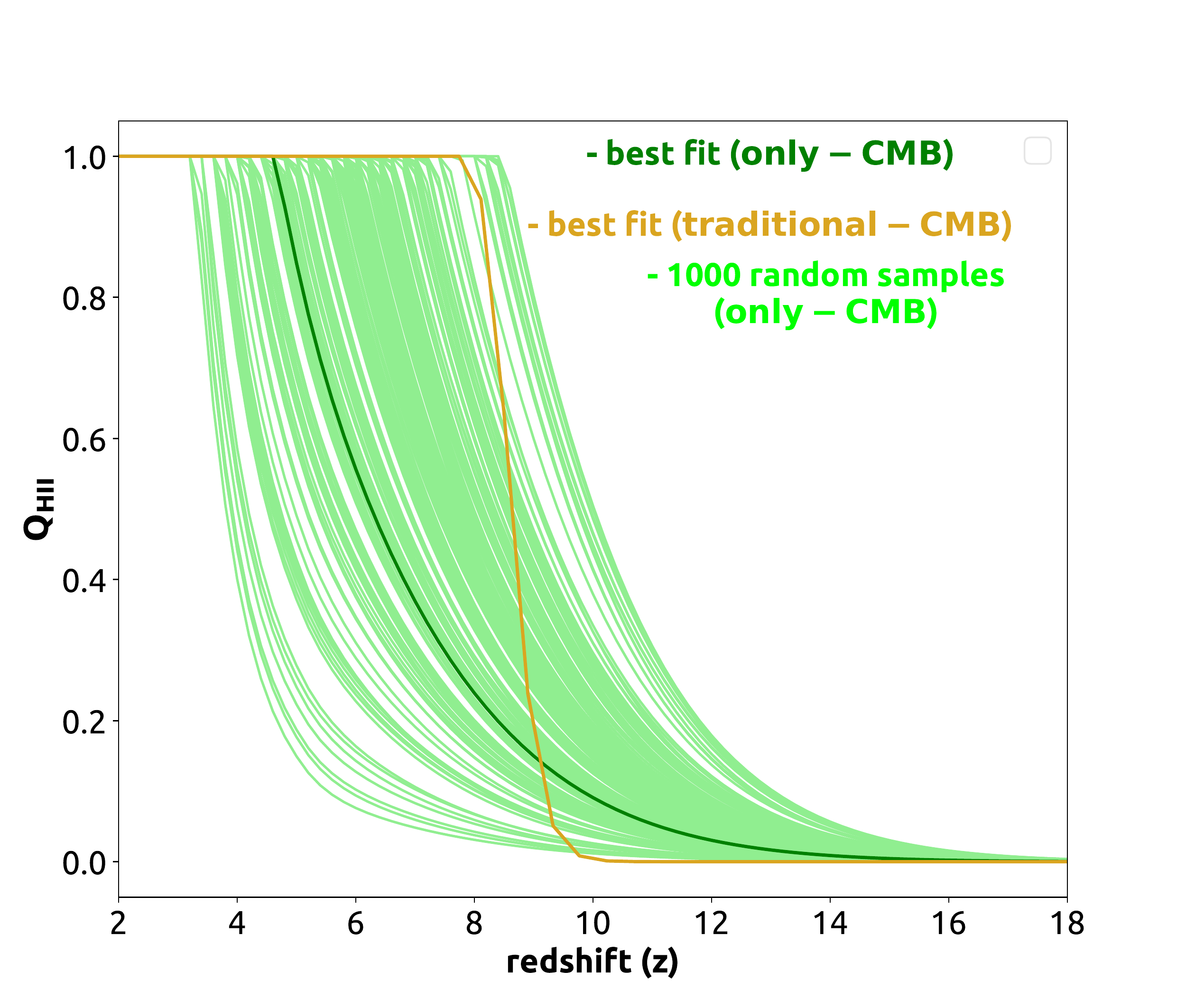}
    \caption{redshift evolution of the volume filling factor of HII region ($Q_{\mathrm{HII}}$)}
    \label{fig:redshift_evolution_only_CMB}
\end{figure}

In this appendix, we show some comparisons of our model with the traditional methods of constraining parameters using the CMB data. For the \textbf{traditional-CMB} analysis, we assume the reionization history to have a tanh form \citep{2020A&A...641A...6P}. The results are then compared with the \textbf{only-CMB} analysis of Section \ref{section:only_CMB} where we use only the CMB data to constrain the parameters using the CF reionization model. The posterior distributions of the model parameters for the two cases are shown in \fig{fig:comparing_tanh_CF}.

As is evident from this figure, the constraints on cosmological parameters are not much different between the two analyses, except for $\sigma_8$ and $\tau$. The mean value of $\tau$ comes out to be 0.058 for the \textbf{only-CMB} analysis which is somewhat higher than that obtained from the \textbf{traditional-CMB} analysis ($\tau=0.052$). Also, the 95\% upper limit of $\tau$ is $0.08$ for the \textbf{only-CMB} is higher than that obtained from the \textbf{traditional-CMB} analysis $(0.07)$. To investigate this in more detail, we have plotted reionization histories coming from the \textbf{only-CMB} analysis with the \textbf{traditional-CMB} in \fig{fig:redshift_evolution_only_CMB}. The light-green shaded region shows the evolution of the volume fraction of ionized Hydrogen ($Q_{\mathrm{HII}}$) corresponding to 1000 random samples obtained from the MCMC chains of the \textbf{only-CMB} analysis. The green curve corresponds to the best-fit values of the free parameters of this analysis. The orange curve corresponds to best-fit values obtained from the \textbf{traditional-CMB} analysis. It is very evident that the allowed CF reionization histories are very different from that of the tanh model. In the CF models, the reionization process starts early (around $z=10$) but extends to much later redshift showing a much slower evolution with redshift compared to the tanh reionization. This extended tail in CF reionization explains the higher value of $\tau$ compared to the traditional Planck estimate. Because $\sigma_8$ and $\tau$ are strongly correlated, higher values of $\tau$ will induce higher values for $\sigma_8$ as well. 

The log-likelihood ($ \mathcal{L}_{\mathrm{Pl}}$) of the best-fit model for the \textbf{only-CMB} analysis  turns out to be $\sim 694$ which is marginally less than that coming from \textbf{traditional-CMB} analysis ($\mathcal{L}_{\mathrm{Pl}} \sim 696$), suggesting that the CF reionization model is an equally good (if not better) match to the data as compared to the traditional tanh model.


\bsp	
\label{lastpage}
\end{document}